\documentclass[pr1,twocolumn,superscriptaddress,showpacs,amsmath,amssymb]{revtex4-1}
\usepackage{graphicx}
\usepackage{indentfirst}
\usepackage{psfrag}
\usepackage{epsfig}
\usepackage{xcolor}
\usepackage{amsmath}
\usepackage{amssymb}
\usepackage{bm}
\usepackage{mathrsfs}
\newcommand{\be}{\begin{eqnarray}}
\newcommand{\ee}{\end{eqnarray}}

\newcommand{\hel}{\mathscr{ H}}

\newcommand{\qui}{\mathscr{ H_R}}

\begin{document}
\title{Reactive helicity and  reactive power in nanoscale optics:  Evanescent waves. Kerker conditions. Optical theorems and reactive dichroism}    
\author{Manuel Nieto-Vesperinas}
\affiliation{Instituto de Ciencia de Materiales de Madrid, Consejo Superior de
Investigaciones Cient\'{i}ficas.\\
 Campus de Cantoblanco, Madrid 28049, Spain. www.icmm.csic.es/mnv}
\author{Xiaohao Xu}
\affiliation{ Institute of Nanophotonics, Jinan University, Guangzhou 511443, China }
\email{mnieto@icmm.csic.es , xuxhao\_dakuren@163.com}
\begin{abstract}
Considering  time-harmonic optical fields,  we put forward the  complex helicity and its  alternating  flow, together with their conservation equation: the complex helicity theorem.  Its imaginary part constitutes a novel  law that rules  the build-up of what we establish as the  reactive helicity through  its zero time-average  flow. Its associated reactive flow, and the imaginary Poynting momentum that accounts for the accretion of reactive power, are illustrated in two paradigmatic systems: evanescent waves and  fields scattered from magnetodielectric dipolar nanoparticles. As for the former, we show that its reactive helicity may be experimentally observed as we \color{red}  introduce   a {\it reactive spin momentum} and a {\it  reactive orbital momentum} in terms of which we express the imaginary field momentum,  whose transversal component produces  an optical force on a magnetoelectric particle that, as we illustrate,  may surpass and  can be discriminated from, the known force due to the so-called extraordinary momentum. We also uncover a non-conservative force on such a magnetoelectric particle, acting in the decay direction of the evanescent wave, and that may also be discriminated from the standard gradient force; thus making the reactive power of the wavefield also observable.
\color{black}
Concerning the light scattered by magnetoelectric nanoparticles, we establish two optical theorems that govern the accretion of reactive helicity and reactive power on extinction of  incident wave helicity and energy. Like a nule total - i.e. internal plus external - reactive power is at the root of a resonant scattered power, we show that a zero total reactive helicity  underlies a resonant scattered helicity. These reactive quantities are shown to yield a novel interpretation of the two Kerker conditions which we demonstrate to be linked to an absence, or minimum, of  the overall scattered reactive energy. Further, the first Kerker condition, under which the particle becomes dual on illumination with circularly polarized light,  we demonstrate to amount to a nule overall  scattered reactive helicity. Therefore, these two  reactive quantities are shown to underly the directivity of the particle  scattering and emission.   In addition, we discover  a discriminatory property of the reactive helicity of chiral light incident on a chiral nanoparticle  by excitation of the  external reactive  power. This  should be   useful for optical near-field enantiomeric separation, an effect that we call reactive dichroism.
\end{abstract}
\maketitle

\section{\label{sec:intro} Introduction} \label{1}
Reactive  quantities of electromagnetic fields, such as reactive and stored energy or the imaginary Poynting vector (IPV), are associated to physical entities that do not propagate in the environment,  like evanescent waves in their varied forms as surface waves \cite{raether,kolokolov,harrington}, standing waves \cite{harrington}, along with near-fields of RF-antennas  \cite{harrington,stratton,wheeler,chu,collin,jackson,mcLean,harrington,balanis,alu}. In recent years, advances in photonics and nano-optics led to  developing concepts such as chirality, helicity \cite{tang,barnett2,corbato,nietoheli,banzer,barnett1,yan}, magnetoelectric effects associated with   the imaginary Poynting vector  (IPV) \cite{nieto1}, and its consequent transverse spin momentum and magnetoelectric energy density \cite{bliokh1,bliokh_rep,bliokh2}, as well as related to the azimuthal imaginary Poynting momentum \cite{xu} and the Kerker-type non-conservative intensity gradient force  \cite{xiao}.

 In spite of progress in the analysis of optical antennas, mainly addressing quantum emitters and plasmonic nanoparticles \cite{novotny1,norris1,barnes}, whose radiative and feeding characteristics  is studied from the point of view of RF-antennas and ciecuit theory \cite{hecht,ziolkowski,engheta}, we have found few detailed studies \cite{ziolkowski} on their reactive quantities; although the effects of reactive power in antenna  functionality  are well-known \cite{harrington,wheeler,chu,collin,mcLean,harrington,geyi,balanis,alu,ziolkowski}; e.g. the radiative and reactive energies of an oscillating dipole  are intertwined. In this way, a major task in RF-antenna design has been the study of its  reactive power and $Q$-factor, seeking a minimization of both quantities in order to match the input energy with its radiative performance,  since a large $Q$ and reactive energy in the antenna  convey  high ohmic losses, and a decrease of its operative bandwidth.

By contrast, at optical wavelengths a high reactive power outside the nanoemitter or scatterer, external stored energy, and $Q$-factor  of a  low-loss nanoparticle, or  nanoantenna, enhances both its scattered (or radiated) power and  frequency sensitivity, also narrowing its operational bandwith, this being desirable for its role as e.g. a nano-source or a biosensor, as well as to reinforce light-matter interactions at the nanoscale, both in plasmonics \cite{saad} and Mie-tronics \cite{won,bonod}. The reason is that, as we shall see, although  such a large external reactive power and stored energy occur at wavelengths near those of resonant scattered power, the interior of the particle acts as a compensating (capacitive or inductive) element  so that its reactive power and stored energy cancel out the external ones close to these resonant wavelengths at which each of these quantities have near extreme values.  The result  is that, in analogy with  RF-antenna design \cite{ziolkowski}, the total (i.e. internal plus external) reactive power and stored energy are zero close to resonances. 

We also show that the same effect occurs for the  interior, external, and total reactive helicities in connection with a maximum efficiency in the helicity scattered  up to the far zone.  

Since surface  plasmons work at optical frequencies where metals exhibit high losses, there has been an increasing interest in high refractive index nanoantennas, on which light exerts a magnetoelectric response  with large electric and magnetic resonances \cite{won,bonod,nietoSi,nieto2011,mlight,staude,kivshar,nietolibrev}.  We will study the key role of reactive quantities  in the scattering from these magnetodielectric particles. The other  archetypical configuration in which we shall address these reactive quantities is an evanescent wave.

Using time-harmonic wavefields, we will start addressing  the  flow of reactive power of  the complex Poynting theorem, and  since an analogous  law for  the helicity flow has never been established, as far as we know, we will introduce the concepts of {\it complex helicity density} \cite{comment0} and   {\it complex helicity flow}, for which we put forward a conservation equation that we coin as {\it complex helicity theorem}.  Its real part is the well-known continuity equation for the conservation of optical helicity \cite{tang,barnett2,corbato,nietoheli,barnett1}, while its imaginary part is a novel law that describes the build-up of {\it reactive helicity}, whose  flow  has zero time-average and hence it does not propagate in free-space. 

We show that this reactive helicity density and its flow  exist, like the reactive energy and the imaginary Poynting vector, in  wavefields that do not propagate into the environment, as e.g. evanescent \color{red} and elliptically (and circularly in particular) polarized standing waves;  \color{black} the former being identical to the so-called magnetoelectric energy density,  introduced in \cite{bliokh2} following symmetry arguments but without providing  its undelying physical law.

As for an evanescent wave, whose  imaginary Poynting vector and its associated transversal spin have been  studied \cite{bliokh1}, while its  time-averaged energy flow, spin angular momentum,  Belinfante momentum, and orbital momentum  have to do with the time-averaged energy and helicity densities \cite{bliokh1}, we  show that the reactive (i.e. imaginary)  Poynting vector, and reactive helicity flow, are linked with the reactive energy and reactive helicity of the wave.   These relationships {\it provide a physical law for our introduced concept of reactive helicity, in which the aforementioned magnetoelectric energy} \cite{bliokh2}, {\it and the  so-called  "real helicity" discussed in a different research} \cite{kamenetskii}, are unified.  Further, we discuss how the reactive power and reactive helicity of this wavefield, are generated close to the interface through their corresponding alternate flows along the wave decay direction.

Moreover, we show that the imaginary  Poynting momentum of the evanescent wavefield is \color{red} the sum of densities of  a {\it reactive spin momentum} and a {\it  reactive orbital momentum}, which in turn are expressed as differences of the imaginary magnetic and electric  corresponding momenta. This leads us  to uncover {\it a non-conservative optical force on a magnetodielectric particle, directed along the  wave decay direction -i.e. different from the gradient  force and larger than it  at certain wavelengths, thus being experimentally detectable- }  {\it due to  the IPV}. 

We also illustrate  the presence of {\it  a transversal force on such a magnetodielectric particle, stemming from the corresponding component of the IPV,  which may be  of opposite sign and much larger  than the known lateral force } \cite{bliokh1} {\it due to the time-averaged Poynting vector; thus being detectable and making  observable the reactive helicity}. \color{black} 

As regards   fields emitted or scattered by a magnetodielectric dipolar nanoparticle, we shall show that its reactive energy and angular distribution of scattered radiation, being  intimately interrelated,  provide a novel interpretation of the two Kerker conditions: K1 and K2 \cite{kerker,nieto2011,nietoJNano,geffrin,lapin,banzer1,olmos}, and hence of the particle emission directivity.  Namely, {\it  under plane wave illumination, and at wavelengths where such a particle  fulfils either  K1 of zero backscattering, or K2 of minimum forward scattering, the overall external reactive power around this body is either zero or almost zero, while the internal and total (i.e. external plus internal) reactive powers are near zero. Moreover,  the overall external reactive helicity vanishes at K1 wavelengths on illumination with circularly polarized light.} 

Concerning feeding  the magnetodielectric nanoantenna, we put forward the  {\it reactive power} and the {\it reactive helicity optical theorems} which quantify the  accretion of external stored reactive  energy and reactive helicity  in the near and intermediate-field zones, in terms of the particle excitation and extinction of  energy and helicity of the supplied illumination.   \color{red}The    effects here shown maximizing both the  external and internal values of these reactive quantities, (while minimizing their overall amounts, i,.e. external plus internal), which tune to resonance  their radiation efficiency,   establishes an analogy with the same well-known pursuit in RF antenna design. Therefore, this work puts forward the importance of reactive quantities  which  underly some previously studied concepts in the analysis of optical antennas \cite{novotny1,norris1,barnes,hecht}. \color{black}

Finally, we show that the reactive power theorem yields an interpretation of how   by illuminating a chiral particle with a non-free propagating wavefield, like an elliptically polarized standing wave, evanescent wave, or near field from a nearby emitter, the reactive helicity of the incident  wavefield appears  as a consequence of the generation of  reactive energy on interaction with the sample  particle. It is intriguing that, as we find, {\it  this incident reactive helicity emerges analogously as the incident  helicity does  in a standard dichroism far-field observation} \cite{tang}. Therefore, we show  that the so-called  "magnetoelectric response"  of a   chiral particle, quoted in \cite{bliokh2},   arises as a  consequence of the accretion of its reactive energy    from near-field chiral illumination, e.g. with incident evanescent waves, (or, similarly, with incident circularly polarized standing waves as proposed in \cite{bliokh2}). As such, we name {\it reactive dichroism} this magnetoelectric phenomenon. It underlines the observability of such incident reactive helicity, and its discriminatory property for enantiomer separation by near-field optical techniques using structured illumination.

 \section{The reactive Poynting vector and the reactive power}\label{sec2}

  To fix some concepts to deal with, we first outline the main quantities involved in the complex Poynting theorem.  
\color{red}
We shall assume an arbitrary body immersed in a lossless homogeneous medium \cite{note1}.
\color{black}
Let  $\bm {\mathcal E}({\bf r}, t)={\bf E}({\bf r})\exp(-i\omega t)$, $\bm{\mathcal B}({\bf r}, t)={\bf B}({\bf r})\exp(-i\omega t)$ be a time-harmonic electromagnetic field. It is well-known  that in a body with charges and free electric currents of density ${\bf j}$  contained in a volume $V$ with permitivity $\epsilon$ and permeability $\mu$,  the complex work density, given by the scalar product ${\bf j}^*\cdot {\bf E}$,  leads after using Maxwell's equations, $\nabla\times \bm{\mathcal E}=-(1/c)\,\partial  \bm{\mathcal B}/\partial t$,  $\,\,$ $\nabla\times  \bm{\mathcal H}=(4\pi/c)\,\bm{\mathcal  J}+(1/c)\,\partial \bm{\mathcal D}/\partial t$,\, $[\bm{\mathcal D}=\epsilon\bm{\mathcal E}, \bm{\mathcal B}=\mu\bm{\mathcal H}, \bm{\mathcal J}({\bf r}, t)={\bf j}({\bf r})\exp(-i\omega t)$],  to the  complex Poynting theorem  \cite{harrington,stratton,jackson,balanis}:
 \be
\int_{V}[\frac{1}{2}\, {\bf j}^* \cdot{\bf E}+ \nabla \cdot {\bf S}]d^3 r = i2\omega\int_{V} (<w_m> -<w_e>) d^3 r\, ; \nonumber \\
  (\omega=\frac{kc}{\sqrt{\epsilon\mu}}=\frac{2\pi}{\lambda}\frac{c}{\sqrt{\epsilon\mu}}).\,\,\, \,\,\,\,\,\,\,\,\, \,\,\,\,\,\,\,\,\, \,\,\,\,\,\,
 \label{bcpoy1}
\ee
Where * and $<.>$ mean complex conjugated and time-average, respectively. The {\it complex   Poynting vector} (CPV) ${\bf S}$, and time-averaged electric and magnetic energy densities, $<w_e>$ and $<w_m>$, are
\be
{\bf S}({\bf r})=\frac{c}{8\pi\mu} {\bf E}({\bf r})\times {\bf B}^*({\bf r}), \, \,\,\,\,\,\,\,\,\, \,\,\,\,\,\,\,\,\, \,\,\,\,\,\, \nonumber \\
<w_e({\bf r})>=\frac{\epsilon}{16\pi}| {\bf E}({\bf r})|^2,\,\, \,\,
<w_m({\bf r})>=\frac{1}{16\pi\mu}| {\bf B}({\bf r})|^2 \,.  \,\,\,\, \,\,\,\, \,\,\,\,\label{bcpoy2}
\ee
Under our above assumptions,  the right side of (\ref{bcpoy1}) is purely imaginary, and the real part of  this equation constitutes the well-known Poynting theorem describing the variation of energy in the body due to the work rate of the field upon its charges,  given by the volume integral of $\frac{1}{2}Re\{ {\bf j}^* \cdot{\bf E}\}$. This variation is characterized by the flow of the time-averaged Poynting vector, $<{\bf S}>=Re\{{\bf S}\}$, across the surface $\partial V$ of $V$: $\int_{\partial V}Re\{ {\bf S}\}\cdot \hat{\bf r}d^2 r$. Where $\hat{\bf r}$ is the outward unit normal to $\partial V$.

On the other hand, the imaginary part of (\ref{bcpoy1}) formulates that  in  the steady state the  integral of the {\it imaginary work}   $\frac{1}{2}Im\{ {\bf j}^* \cdot{\bf E}\}$, plus  the {\it reactive power flux} $\int_{\partial V}Im\{ {\bf S}\}\cdot{\bf n}d^2 r$, (which has zero time-average since it is associated with instantaneous energy flow that alternates back and forth at twice frequency  across  $\partial V$), accounts for this  {\it  reactive  power} build-up in and around the body, given by the right side of  (\ref{bcpoy1}); $2(<w_m> -<w_e>)$ being the {\it reactive energy density}. Unless otherwise stated, we shall henceforth assume $\epsilon=\mu=1$ for the embedding space. 

At this point it is convenient to introduce the  instantaneous  Poynting vector \cite{harrington, balanis} built by the   fields ${\cal  E}({\bf r},t)=\Re\{\bm {\mathcal E}({\bf r},t)\}$ and  ${\cal  B}({\bf r},t)=\Re\{\bm {\mathcal B}({\bf r},t)\}$,
\be
\bm{\mathcal S}({\bf r},t)=\frac{c}{4\pi}{\cal  E}({\bf r},t)\times {\cal  B}({\bf r},t)=\nonumber \,\,\,\, \,\,\\
<{\bf S}>+ \frac{c}{8\pi}Re\{ {\bf E}({\bf r})\times {\bf B}({\bf r})\exp(-2i\omega t)\}, \nonumber\,\,\,\, \,\,
\ee
This is the standard expression of  $\bm{\mathcal S}({\bf r},t)$. However we find it more instructive to  write  it as:
\be
\bm{\mathcal S}({\bf r},t)=
\,<{\bf S}>(1+\cos 2\omega t)+Im\{{\bf S}\} \sin2\omega t
\nonumber  \,\,\,\, \,\, \,\,\,\, \,\,   \,\,\,\, \,\,    \\
+ \frac{c}{4\pi} {\bf E}^R({\bf r})\times {\bf B}^I({\bf r})\sin2\omega t
- \frac{c}{4\pi} {\bf E}^I({\bf r})\times {\bf B}^I({\bf r})\cos2\omega t. \,\,\,\,\,\ \,\,\,\label{geninst}. 
\ee
Where the superscripts $R$ and $I$ denote real and imaginary parts, respectively.
Note that while the term with   $<{\bf S}>$ does not change sign, as expected from that part of the instantaneous Poynting vector associated with the time-averaged energy flow,  the term that contains the IPV or {\it reactive Poynting vector}, $Im\{{\bf S}\}$,   alternates its sign  at frequency $2\omega$ following the variation of $\sin2\omega t$. This is in accordance with the above interpretation of the imaginary part of (\ref{bcpoy1}). We also see  that there is a generally non-zero contribution to this alternating flow  in the last two terms of (\ref{geninst}). Obviously only the $<{\bf S}>$ term remains on time-averaging in  (\ref{geninst}).

\subsection{The reactive Poynting vector and  the angular spectrum of plane waves}
To distinguish   the structure of the real and imaginary parts of ${\bf S}$, it helps to employ the angular spectrum representation of the electromagnetic field, so as to map these quantities into their spectra in Fourier space. To this end,
we calculate the flow of    $\frac{ c}{8\pi} {\bf E}({\bf r})\times {\bf B}^*({\bf r})$ across a  plane $z=constant$.  We use for simplicity a framework such that  the souces are on $z< 0$ and thus the integration is done on the $z=0$ plane. But one may equally choose any other constant value $z=z_0$, providing the sources lie in $z<z_0$. The electric field  ${\bf E}({\bf r})$ propagating into the half-space $z\geq 0$ is represented by its angular spectrum  of plane wave components  as \cite{nietolibro,mandel}:
\be
{\bf E}({\bf r})=\int_{-\infty}^{\infty}d^2{\bf K}\,\, {\bf e}({\bf K}) \exp [i( {\bf K}\cdot {\bf R}+ k_z z)].  \label{bcfor19}
\ee
  ${\bf r}=({\bf R}, z)$, ${\bf R}=(x,y)$, 

${\bf k}=({\bf K}, k_z)$,  ${\bf K}=(K_x,K_y)$,   $|{\bf k}|^2=k^2$.  And
\be
k_z=\sqrt{k^2- K^2}=q_h \, , \,K\leq k,\,\, \mbox{for homogeneous }\nonumber \\
\mbox{(propagating) plane wave components. } \nonumber\\
k_z=i\sqrt{K^2- k^2}=i q_e \, , \,K> k, \,\, \mbox{for evanescent  }   \nonumber \\
\mbox{plane wave components. } \,\,\,\,\,\,\,\, \,\,\,\,\,\,\,\,  \label{bcfor19bis}
\ee 
 
 Using the subscripts $h$ and $e$ for homogeneous and  \color{red}evanescent \color{black} components, respectively, the CPV flux,  $ \Phi^{Poynt}$, across the plane $z=0$ has real and imaginary parts given by, (see the proof in Appendix A):

\be
\Phi^{RPoynt}= \frac{\pi c}{2k}\int_{K\leq k}d^2{\bf K}\,q_h |{\bf e}_h({\bf K})|^2,
 \label{bcfor26a1}
\ee
and
\be
\Phi^{IPoynt}=-\frac{\pi c}{2k}\int_{K> k}d^2{\bf K}\,[q_e|{\bf e}_e({\bf K})|^2 -2q_e|{ e}_{e\,z } ({\bf K})|^2]. \,\,\,\,\,\,\,\,\label{bcfor26b1}
\ee
Eq.(\ref{bcfor26a1}) is well-known, it expresses  the flux $\Phi^{RPoynt}$ of the real part of the CPV {  as the momentum power carried on by  the propagating components}, ($K\leq k$). However, Eq. (\ref{bcfor26b1})  shows that {\it the flux $\Phi^{IPoynt}$ of the reactive CPV {\it is  momentum associated with power contained in  the evanescent components}, $(K> k)$, and as such it does not propagate into $z\geq 0$}. Notice the special role played by the longitudinal component ${ e}_{e\,z } ({\bf K})$ in (\ref{bcfor26b1}).

\section{The reactive helicity and its  reactive flow theorem}
The optical {\it helicity} density of the electromagnetic field in a medium with constitutive parameters $\epsilon$ and $\mu$ is  well-known to be \cite{barnett2,nietoheli,barnett1}
\be
{\hel}({\bf r})=(1/2k)\sqrt{\frac{\epsilon}{\mu}}Im\{ {\bf E}({\bf r})\cdot{\bf B}^*({\bf r})\}.\,\, \,\,\,\,\,\,\,\,\,\,\,\,\,\,\,\label{heli}
\ee
 We now introduce the quantity\color{red}
 \be 
{\qui}({\bf r}) =(1/2k)\sqrt{\frac{\epsilon}{\mu}}Re\{ {\bf E}({\bf r})\cdot{\bf B}^*({\bf r})\}.  \,\,\,\,\,\,\,\,\,\,\,\,\,\,\,\, \label{reactheli}
\ee
\color{black}
 Like $Im\{{\bf S}\}$, this quantity $\qui$ may  appear from ${\hel}$ when ${\bf E}$ and ${\bf B}$ are chosen $\pi/2$ out of phase. 
 
  We shall later see  that  $\qui$ {\it is exclusive of}  $Im\{{\bf S}\}$, (and of course of a reactive helicity flow, to be introduced  next).  Thus we call  ${\qui}$  the {\it reactive helicity} density of the field.  Some authors have recently addressed this quantity in different works, and call it magnetoelectric energy \cite{bliokh2}, or just  real helicity \cite{kamenetskii}. However we keep our denomination by showing that, like the reactive energy, it fulfills a conservation law.

 We consider a body with charges and free currents, embedded in a volume $V$. From  the two Maxwell  equations for the spatial vectors, $\nabla\times {\bf E}=ik{\bf B}$ and $\nabla\times  {\bf H}=(4\pi/c)\,{\bf j} -ik{\bf D}$,  we  derive the conservation equation for the reactive helicity.  First, we employ the second of these equations and address the  scalar product $(4\pi/c){\bf j}^*\cdot {\bf B}$ in $V$. On using the identity: $ {\bf B} \cdot (\nabla \times  {\bf H^*})={\bf H}^* \cdot (\nabla \times  {\bf B})- \nabla\cdot ({\bf B} \times  {\bf H}^*)$, one gets 
  \be
\int_{V}[-\frac{2\pi }{kn} Im{\{\bf j}^* \cdot{\bf B}\}+ \, \nabla \cdot  \bm{\mathcal F}_B] \, d^3 r 
=\omega\int_{V}  \,d^3 r \,{\qui} .  \,\,\,\, \,\,\,\,\, .  \label{bcpoy33}
\ee
In (\ref{bcpoy33}) $\bm{\mathcal F}_B=(c/4kn)\Im\{{\bf H}^* \times {\bf B}\}$ is the density of  magnetic flow of helicity  \cite{barnett2,nietoheli,bliokh2}, ($n=\sqrt{\epsilon\mu}$);  and it is also proportional to the magnetic part of the spin angular momentum density \cite{barnett2,nietoheli,bliokh2}, (see also Eq. (\ref{spinevan}) below).

Similarly,  we may obtain a conservation equation for   $\qui$ with the electric helicity flow density, $\bm{\mathcal F}_E=(c/4kn)Im\{{\bf E}^* \times {\bf D}\}$. This is done by taking the scalar product:
${\bf D}\cdot  (\nabla\times {\bf E}^*)=-ik\,{\bf D}\cdot {\bf B}^*$ ,  and proceeding in an identical way as with the derivation of   (\ref{bcpoy33}), the result is
  \be
\int_{V} \nabla \cdot \bm{\mathcal F}_E \, d^3 r = - \omega\int_{V} \qui \,d^3 r \,. \,\,\,\,\,\,\,\,\,\,\,\,\,\,\,\,\,\, \label{bcpoy43}
\ee
Adding (\ref{bcpoy43}) and (\ref{bcpoy33}) one obtains the well-known continuity equation for the conservation of  helicity in the steady state:
\be
\int_{V}\nabla \cdot\bm{\mathcal F}\, d^3 r\equiv\int_{\partial V}\bm{\mathcal F}\, \cdot \hat{\bf r}\,d^2 r=
\frac{2\pi}{kn}\int_{V} Im{\{\bf j}^* \cdot{\bf B}\} \, d^3 r  , \,\,\,\,\, \,\,\,\,\,  \, \,  \label{bcpoyReal}
\ee
which shows that  the flow of helicity, or dual-symmetric spin \cite{bliokh4,nietoheli}:  
\be
\bm{\mathcal F}=\bm{\mathcal F}_E + \bm{\mathcal F}_B=(c/4kn)Im\{{\bf E}^* \times {\bf D}+{\bf H}^* \times {\bf B}\} , \,\,\,\,\,\label{helflow}
\ee
 across the boundary $\partial V$  of $V$ equals the radiated field helicity, including  its dissipation and conversion  given by the right side of (\ref{bcpoyReal}) \color{red}  \cite{poulikakos1,poulikakos2,gutsche}.\color{black}

However, substracting (\ref{bcpoy43}) from (\ref{bcpoy33}) leads to the {\it reactive helicity flow theorem}, 
\be
\int_{V}[-\frac{2\pi }{kn} Im{\{\bf j}^* \cdot{\bf B}\}+ \nabla \cdot \bm{\mathcal F}_{{\qui}}] \, d^3 r = 2 \omega \int_{V} \qui \,d^3 r \, ;   \,\,\,\,\,\,\,\,\,\,  \,\,\,\,\, \label{bcpoy53}
\ee
where we have introduced the {\it reactive helicity flow} \color{red} associated to a flow of helicity that vanishes on time-average, although not  instantly. \color{black}

\be
 \bm{\mathcal F}_{{\qui}}=\bm{\mathcal F}_B - \bm{\mathcal F}_E=  
(c/4kn)\Im\{{\bf H}^* \times {\bf B}-{\bf E}^* \times {\bf D}\}= \nonumber \\
 \frac{2\pi c^2}{n}(<\bm{\mathcal S}_m>-<\bm{\mathcal S}_e>). \,\,\,\,\,\,\,\,\, \label{reactflow}
\ee 
If  $\partial V$ is in air or vacuum,  $n=1$, ${\bf B}={\bf H}$ and ${\bf D}={\bf E}$ in (\ref{helflow}) and  (\ref{reactflow}). $<\bm{\mathcal S}_e>$ and $<\bm{\mathcal S}_m>$ stand for the time-averages of the density of electric and magnetic spin angular momentum, (cf. Appendix C).

The quantity $2\omega \qui$ is a reactive helicity per unit half-period, \color{red} or just the {\it  reactive helicity power}, \color{black} in analogy with the reactive power of the complex Poynting theorem Eq. (\ref{bcpoy1}). \color{red}Hence (\ref{bcpoy53}) expresses the conservation of $2\omega \qui$\color{black}.  Indeed  (\ref{bcpoyReal}) and (\ref{bcpoy53})  \color {red}suggest us to formulate
 a {\it complex helicity  theorem} \cite{norris_use}: \color{black}
\be
\int_{V}[-(1+i)\frac{2\pi }{kn} Im{\{\bf j}^* \cdot{\bf B}\}+ \,\nabla \cdot \bm{\mathcal F}_C] \, d^3 r = 2i \omega\int_{V} \qui \,d^3 r ,
\nonumber \\
 \bm{\mathcal F}_C =\bm{\mathcal F} +i \bm{\mathcal F}_{{\qui}}. \,\, \,\,\,\,\,  \,\, \,\,\,\,\, \,\,\,\,\,  \,\, \,\,\, \label{bcpoy63}
\ee
where the {\it complex helicity flow} , or {\it complex spin angular momentum}, is $ \bm{\mathcal F}_C $.  Evidently  (\ref{bcpoy63}) has a real part which is the standard helicity conservation equation (\ref{bcpoyReal}), whereas its imaginary part  is Eq. (\ref{bcpoy53}) for the reactive helicity flow and   governs the variation of reactive helicity in $V$, given by  the decrease of the integrated source density $-\frac{2\pi}{kn} Im{\{\bf j}^* \cdot{\bf B}\}$. This variation is expressed in terms of the {\it reactive} helicity flux $\int_{\partial V} \bm{\mathcal F}_{{\qui}}\cdot\hat{\bf r}\,d^2 r$, (which has zero time-average since it comes from $\Im\{ \bm{\mathcal F}_C\}$, and hence represents helicity flowing back and forth to the body across  $\partial V$ in its near-field region, without net propagation), and of integrated  {\it  reactive  helicity density}  $2\omega   \qui$.   

Unless otherwise stated, we shall not drag the factor $2\omega$ when we refer to the reactive helicity, thus we shall just write $\qui$ for this quantity. In Section VII we show that $\qui$ is built-up on chiral light-matter interaction. This gives rise to  the phenomenon of reactive dichroism in the near-field of the body,  \color{red}  addressed in Section VIII. \color{black}

\color{red}
Equations (\ref{bcpoy33}) and (\ref{bcpoy43}) suggest that $\qui$ is observable, for example by detecting the torque exerted by a circularly polarized plane wave on a dipolar particle on which the field induces a purely electric (e) or a purely magnetic (m) dipole.  In fact, in vacuum $\bm{\mathcal F}_E$ and $\bm{\mathcal F}_B$ are proportional to the optical electric  and magnetic  torque, respectively,   $\bm{\Gamma}_E=(1/8\pi k)\sigma_e^{(a)}Im \{{\bf E}^*\times{\bf E}\}=c\sigma_e^{(a)}<\bm{\mathcal S}_e>$ and $\bm{\Gamma}_B=(1/8\pi k)\sigma_m^{(a)}Im \{{\bf B}^*\times{\bf B}\}=c\sigma_m^{(a)}<\bm{\mathcal S}_m>$; $\sigma_m^{(a)}$ and $\sigma_m^{(a)}$ being the particle electric and magnetic absorption cross sections, (see \cite{nieto_torque}, Section X). Otherwise, if the particle is magnetodielectric, $\qui$ becomes observable through the e-m interaction force, Eq. (\ref{Fem}) below, [cf. \cite{nieto1} Eq. (44)]. This latter situation is detailed with an evanescent wave in  Section IV.B, cf. Eq. (\ref{ImPevan}).
\color{black}

Concerning the $P$, $T$, $D$ {\it symmetries} of these novel quantities, namely,

{\it parity}, $P: {\bf r} \rightarrow -{\bf r}$,  $P: {\bf E} \rightarrow -{\bf E}$,  $P: {\bf B} \rightarrow {\bf B}$, 

 {\it time-reversal}, $T: t \rightarrow -t$,  $T: {\bf E} \rightarrow {\bf E}^*$,  $T: {\bf B} \rightarrow  -{\bf B}^*$,
 
and {\it  duality},  $D: {\bf E} \rightarrow {\bf B}$,  $D: {\bf B} \rightarrow -{\bf E}$,

$\\$
 while it is well-known that
$\\$

 $P: <{\bf S}> \rightarrow - <{\bf S}>$,  $P: \bm{\mathcal F} \rightarrow \bm{\mathcal F}$,  $P: {\hel} \rightarrow -{\hel}$;  

 $T: <{\bf S}> \rightarrow - <{\bf S}>$,  $T: \bm{\mathcal F}\rightarrow  - \bm{\mathcal F}$,  $T: {\hel} \rightarrow {\hel}$;

$D: <{\bf S}> \rightarrow  <{\bf S}>$,  $D: \bm{\mathcal F} \rightarrow \bm{\mathcal F}$,  $D: {\hel} \rightarrow {\hel}$;

$\\$
 and we obtain the symmetries for the reactive quantities:

$\\$

$ P: Im\{{\bf S}\} \rightarrow -Im\{{\bf S}\}$, $ P: \bm{\mathcal F}_{\qui} \rightarrow  \bm{\mathcal F}_{\qui}$, \,\,\,\,\,\,\,\,\,\,\,\,\,\,\,\,\,\,\,\,\,\,\,\,
 $P: {\qui} \rightarrow -{\qui}$; 

 $T: Im\{{\bf S}\} \rightarrow Im\{{\bf S}\}$, $ T: \bm{\mathcal F}_{\qui} \rightarrow -\bm{\mathcal F}_{\qui}$,  \,\,\,\,\,\,\,\,\,\,\,\,\,\,\,\,\,\,\,\,\,\,\,\,
 $T: {\qui} \rightarrow -{\qui}$; 

 $D: Im\{{\bf S}\} \rightarrow - Im\{{\bf S}\}$,   $D: \bm{\mathcal F}_{\qui} \rightarrow -\bm{\mathcal F}_{\qui}$,  \,\,\,\,\,\,\,\,\,\,\,\,\,\,\,\,\,\,\,\,\,\,\,\,
 $D: {\qui} \rightarrow -{\qui}$.

$\\$

It is interesting that while under parity  the three reactive quantities behave like their corresponding non-reactive counterparts, they invert their symmetry under duality applications, in contrast with their  non-reactive analogues that remain $D$-invariant.
Under time-reversal, only $\bm{\mathcal F}_{\qui}$ has the same symmetry as its non-reactive correspondant $\bm{\mathcal F}$.

\subsection{The reactive helicity and the angular spectrum}

In order to gain more insight into the different nature of ${\hel}$ and ${\qui}$, we shall employ  once again the angular spectrum  of the electromagnetic wave. We evaluate the total ${\hel}$ and ${\qui}$ per unit z-length by integration of   $ {\bf E}({\bf r})\cdot{\bf B}^*({\bf r})$ on a  plane $z=constant$.  Again, we choose  coordinates such that  the souces are on $z< 0$ and thus the integration is done on the $z=0$ plane.

Using, as before, the subindex $h$ and $e$ for homogeneous and evanescent components, respectively, the integral on $z=0$ of  ${\bf E}({\bf r})\cdot{\bf B}^*({\bf r})$ is shown  in appendix B to yield
\be
\int_{-\infty}^{\infty} d^2 {\bf R}\,{\hel}({\bf R},0)\nonumber\\  
=(1/2k)\sqrt{\frac{\epsilon}{\mu}}\int_{-\infty}^{\infty} d^2 {\bf R}\,Im\{{\bf E}({\bf R})\cdot{\bf B}^*({\bf R})\}    \nonumber\\
=2\frac{(2\pi)^2}{kc} \{ \int_{K\leq k}d^2{\bf K}\, {\bf k}_h \cdot  \bm{\mathcal F}_{E\,h}({\bf K})  \,\,\,\,\,\,\,\,\,\,\,\,\,\,\,\,\,\,\,\,\,\ \nonumber\\
+ \int_{K> k}d^2{\bf K}\,{\bf K}\cdot  \bm{\mathcal F}_{E\,e\,\perp}({\bf K})\}, \,\,\,\,\,\,\,\,\,\,\,\,\,\,\,\,\,\,\,\,\,\,
 \label{bcfora2011}
\ee
and
\be
\int_{-\infty}^{\infty} d^2 {\bf R}\,{\qui}({\bf R},0)  \nonumber\\
=(1/2k)\sqrt{\frac{\epsilon}{\mu}}\int_{-\infty}^{\infty} d^2 {\bf R}\,Re\{{\bf E}({\bf R})\cdot{\bf B}^*({\bf R})\} \nonumber \\
=2\frac{(2\pi)^2}{kc}
 \int_{K> k}d^2{\bf K}\, q_e \,  {\bf \cal F}_{E \,e\,z}({\bf K}). \,\,\,\,\, \,\,\,\,\,\,\,\,\,\,\,\,\,\,\,\, \label{bcfora2021a}
\ee
In   (\ref{bcfora2011}) $ \bm{\mathcal F}_{E\,e}=( \bm{\mathcal F}_{E\,e\,\perp}, {\bf \cal F}_{E\,e\,z})$,  $\bm{\mathcal F}_{E\,e\,\perp}=({\bf \cal F}_{E \,e\,x}\,,{\bf \cal F}_{E \,e\,y})$.
Therefore  Eq. (\ref{bcfora2011}) shows that  in the domain of propagating components
 the helicity density of the field in $z=0$ maps into the  projection of the electric spin angular momentum of the ${\bf K}$-plane wave component onto the  propagation wavevectors; namely, onto the real wavevector   ${\bf k}_h$, while in the evanescent region, it is given by the projection of the electric spin onto the transversal (propagating) component ${\bf K}$. (Note that this is in agreement with the standard definition \cite{corbato}, but here generalized to include evanescent waves).

 Of special interest  is, however,  Eq. (\ref{bcfora2021a}), which shows that the reactive helicity density ${\qui}$ maps  in ${\bf K}$-space as {\it the projection of the electric spin of  the  ${\bf K}$th-evanescent component   onto the $z$-component}  $i{\bf k}_z^* =i(-iq_e)\hat{\bm z}$.  Where   ${\bf k}$ is the complex wavevector of this  ${\bf K}$th-evanescent wave, [cf. Eq. (\ref{bcfor19bis})]. Again,  this  justifies that we  call {\it reactive} to the real part of  $ {\bf E}({\bf r})\cdot{\bf B}^*({\bf r})$.

 In this connection, it should be remarked that like  the  reactive power and the  IPV are linked with non-propagating waves, e.g. evanescent and  standing waves \cite{harrington}, the reactive helicity ${{\qui}}$ (and hence its flow $\bm{\mathcal F}_{{\qui}}$) exist in evanescent waves as shown in Eq.(\ref{bcfora2021a}), as well as in \color{red}elliptically polarized (and circularly polarized in particular, CPL)  standing wavefields \cite{note2}.  For   CPL waves  \color{black} the authors of  \cite{bliokh2} used the term  "magnetoelectric energy" \color{red}which, as seen above, is a reactive quantity since it is the same as $\qui$. \color{black}

\begin{figure*}[t!]
\begin{centering}
\includegraphics[width=14.0cm]{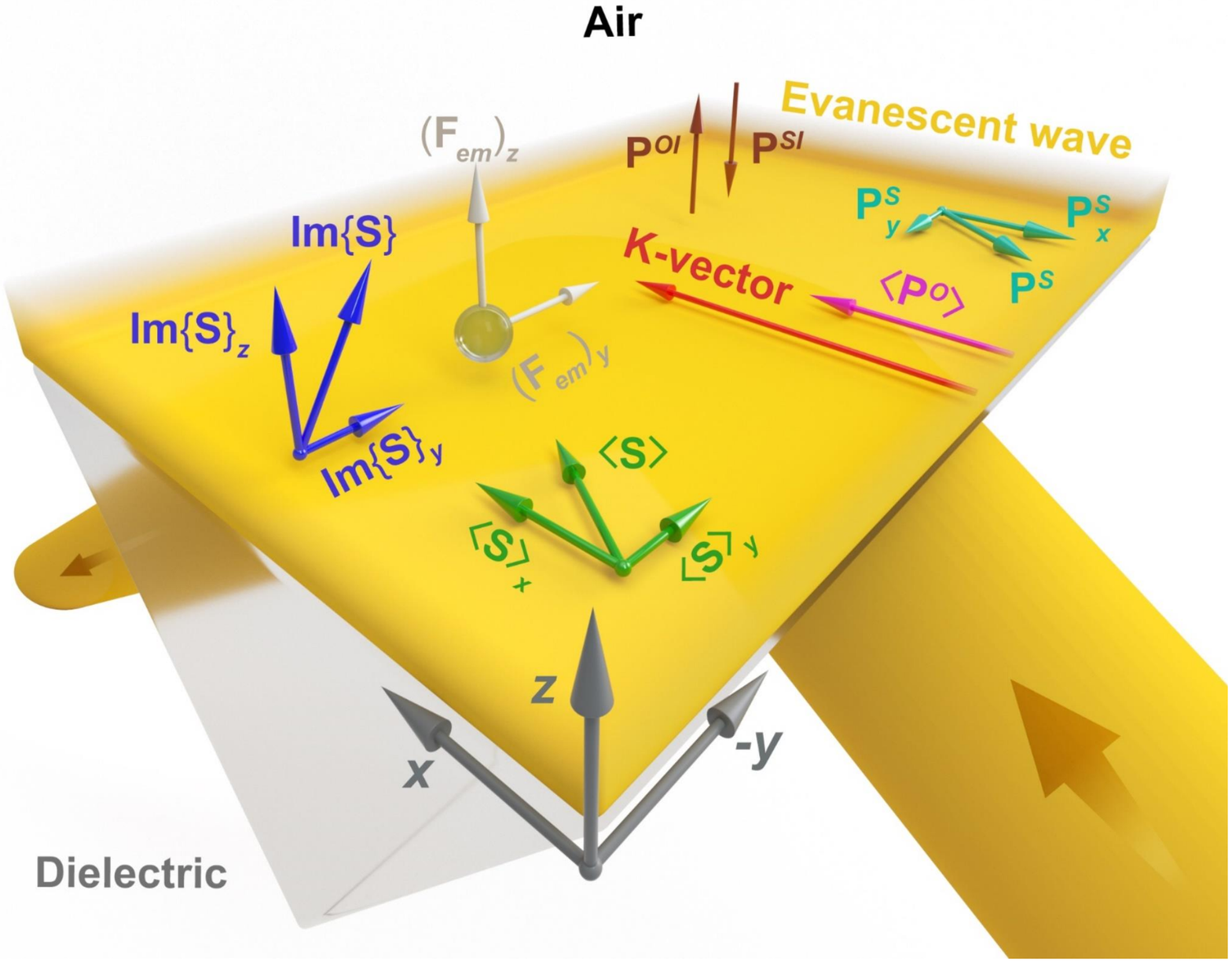}
\par\end{centering}
\caption{The evanescent wave (yellow) of Eq. (\ref{evan}), with reactive power density  given by (\ref{evanreenerg}),
propagates  in the air, parallel to the interface $z=0$ separating it from a dielectric in $z< 0$,  with $K$-vector along $OX$. Its amplitude decays as $\exp(-qz)$. The  $<{\bf S}>$-vector  and  spin momentum, $<{\bf P}^S>$, contained in  $OXY$,  have $y$-components proportional to the wave helicity ${\hel}$, (see Appendix C). The orbital momentum,  $<{\bf P}^0>$, proportional to the energy density $w$ points along $OX$.  The reactive momentum $Im\{{\bf S}\}$, contained in $OYZ$, and whose $z$-component is associated to energy flux bouncing up and down along $OZ$, [cf. Eq.(\ref{divPevan})], may be observed through the time-averaged force ${\bf F}_{em}$, proportional to $Im\{{\bf S}\}$, on a dipolar magnetodielectric particle placed on the interface. This force is due to the interference of the particle electric and magnetic induced dipoles. The lateral $y$-force, $({\bf F}_{em})_y$, and transversal $z$-component, $({\bf F}_{em})_z$ , due to $Im\{{ S}\}_y$ and  $Im\{{ S}\}_z$,  are proportional to the wave reactive helicity, $\qui$, and reactive power density, $\nabla\cdot Im\{{\bf S}\}$, respectively. The imaginary orbital and spin momenta, ${\bf P}^{O\,I}$ and ${\bf P}^{S\,I}$, point along the $+z$ and $-z$-direction, respectively.
The flow $\bm{\mathcal F}_{\qui} $ of reactive helicity is, like $Im\{\bf{ S}\}$, in the $OYZ$-plane with  $y$-component proportional to both the reactive power density, $w_{react}$, and the magnitude of the $K$-vector; while its $z$-component is proportional to the reactive helicity density ${\qui} $, so that across the $OXY$-plane  $({\mathcal F}_{\qui})_z $ holds the imaginary part of the complex helicity theorem, as shown by Eq. (\ref{divReevan}).  }
\end{figure*}

\section{Case 1: Reactive  power, reactive helicity, and reactive momenta in an evanescent wave}
As illustrated in Fig. 1, we  consider a generic time-harmonic evanescent wave in air, $z\geq 0$, generated by   {\it total internal reflection} (TIR),  at a plane interface $z=0$ separating  air ($\epsilon=\mu=1$) from a dielectric in the half-space $z\leq 0$.  The plane of incidence being $OX Z$. Then the complex spatial parts of the electric and magnetic vectors in $z>0$, are expressed in a Cartesian coordinate basis $\{\hat{\bf x},\hat{\bf y},\hat{\bf z}\}$ as ($n=1$)
 \cite{born,nietoOL}:
\be
{\bf E}= \left(-\frac{iq}{k } T_{ \parallel}, T_{\perp},
\frac{K}{k } T_{\parallel}\right) \exp(iKx-qz), \nonumber \\
{\bf B}= n \left(-\frac{iq}{k } T_{ \perp}, -
T_{\parallel}, \frac{K}{k } T_{ \perp}\right) \times \exp(iKx-qz) .\,\,\,\, \label{evan}
\ee 
For TE or $s$ (TM or $p$) - polarization , i.e. ${\bf E}^{(i)}$ (${\bf B}^{(i)}$) perpendicular to the plane of incidence $OXZ$, only those components with the transmission coefficient $T_{ \perp}$,
($T_{\parallel}$) would be chosen in the incident fields \cite{born}. $ K$  denotes the component, parallel to the
interface, of the wavevector ${\bf k}$: $k({\bf s}_{xy}, s_z)= ( K,
0, iq)$, $q= \sqrt{K^2 -k ^2}$, $k ^2= K^2-q^2$.

\color{red}\subsection{Reactive power and reactive helicity densities} \color{black}

 The densities of energy, $w=w_e+w_m$, and reactive power, $w_{react}=2\omega (w_m-w_e)$,  ($\epsilon=\mu=1$), of this wave are according to (\ref{bcpoy2})
\be\,
w=\frac{1}{8\pi}\frac{K^2}{k^2}(|T_{\perp}|^2+|T_{\parallel}|^2)\exp(-2qz), \,\,\,\,\,\,\,\,\,\,\,\ \nonumber \\
w_{react}= \frac{c}{4\pi}\frac{q^2}{k}(|T_{\perp}|^2-|T_{\parallel}|^2)\exp(-2qz). \,\,\,\,\,\,\,\,\,\,\,\,\label{evanreenerg}
\ee
And the densities of  {helicity}, $\hel$, and  reactive helicity, $\qui$,  ($\epsilon=\mu=1$),   of this evanescent wave are
\be
{\hel}=-(K^2/k^3)Im\{T_{\perp}T_{\parallel}^*\}\exp(-2qz); \,\, \nonumber  \\
\qui= (q^2/k^3)Re\{T_{\perp}T_{\parallel}^*\}\exp(-2qz). \,\,\,\, \,\,\,\, \,\,\,\, 
\label{HK}
\ee

\color{red} \subsection{The reactive energy flow: Reactive momentum, imaginary spin and imaginary orbital momenta } \color{black}
The CPV is written as
\be
{\bf S}=\frac{c}{8\pi\mu}[\frac{K}{k}(| T_{ \perp}|^2+|T_{\parallel}|^2), \nonumber \\
 i 2\frac{Kq}{k^2}T_{\perp}^*T_{\parallel}\,, -
 i \frac{q}{k}(| T_{ \perp}|^2-|T_{\parallel}|^2)]   
\exp(-2qz) \,\,\,\,\,\,\,\,\,\,\nonumber \\
=[\frac{ck}{ K} w\,, \frac{c}{4\pi }(-\frac{kq}{K}\hel + i\frac{kK}{q}\qui)\,,-i\frac{1}{2\mu q}w_{react}].\,\,\,\,\,\,\,\,\,\,\,\label{CPevan}
\ee
For the sake of comprehensiveness, in Appendix C we present the  well-known main time-averaged quantities. Here we concentrate on those reactive  less-known and their interralations. 

The reactive, or  imaginary, part of the CPV is
\be
Im\{{\bf S}\}=\frac{c}{8\pi\mu}[\,0\,, 2\frac{Kq}{k^2}Re\{T_{\perp}T_{\parallel}^*\}\,,
 \frac{q}{k}(| T_{ \parallel}|^2-|T_{\perp}|^2)]  \nonumber \\ \times\exp(-2qz) 
= \frac{1}{2q}[0\,,\frac{ckK}{2\pi }{\qui}\,,-w_{react}]
.\,\,\,\,\,\,\, \label{ImPevan}
\ee
Which yields the \color{red}{\it reactive} or \color{black} {\it imaginary  momentum of the field}, (sometimes called imaginary Poynting momentum \cite{bliokh1,xu}), ${\bf g}^I=Im\{{\bf g}\}=Im\{{\bf S}\}/c^2$:

\be
{\bf g}^{I}=\frac{1}{8\pi c}[\,0\,, 2\frac{Kq}{k^2}Re\{T_{\perp}T_{\parallel}^*\}\,,
 \frac{q}{k}(| T_{ \parallel}|^2-|T_{\perp}|^2)]  \nonumber \\ \times\exp(-2qz) 
= \frac{1}{2qc^2}[0\,,\frac{ckK}{2\pi }{\qui}\,,-w_{react}],\,\,\,\,\,\,\,\,\,\,\,\,  \label{ImPevan1}
\ee
\color{red}
whose components come from  two vectors that we put forward next: {\it the density of both reactive spin momentum } $\bm{\mathcal P}^S$ and  {\it reactive orbital momentum} $\bm{\mathcal P}^O$:
\be
\bm{\mathcal P}^S=\frac{1}{2}({\bf P}_m^{S\,I}-{\bf P}_e^{S\,I})=\frac{K}{2cq}(0,\frac{k}{2\pi }\qui,-\frac{K}{ck^2}w_{react}), \nonumber \\
\bm{\mathcal P}^O=\frac{1}{2}({\bf P}_m^{O\,I}-{\bf P}_e^{O\,I})=\frac{q}{2cq}(0,0, \frac{q}{ck^2 }w_{react})
.\,\,\,\,\,\,\,\,\,\,\,\,\,\,\,\,\,\,\,\,\, \,\, \label{Imoms}
\ee
Namely,
\be
{\bf g}^I=\bm{\mathcal P}^S+\bm{\mathcal P}^O.  \,\,\, \label{bcfor5A8}
\ee
The electric and magnetic imaginary spin and orbital momenta of (\ref{Imoms}) are given in Appendix C, Eqs. (C-10)-(C-14).

 \color{red} 
  We emphasize that, as seen in (\ref{ImPevan1})-(\ref{bcfor5A8}),  {\it $g_y^I$  is  fully due to ${\mathcal P}_y^S $, while $g_z^I$ comes from  ${\mathcal P}_z^S+{\mathcal P}_z^O$}.

Note that, interestingly and  in contrast with Eq.(\ref{orbplusspin}), denoting ${\bf P}^{S\,I}=(1/2)({\bf P}_e^{S\,I}+{\bf P}_m^{S\,I}) = (0,0, -\frac{q}{kc}w)$, \,\,${\bf P}^{O\,I}=(1/2)({\bf P}_e^{O\,I}+{\bf P}_m^{O\,I})= (0,0, \frac{q}{kc}w)$,  one obtains: ${\bf P}^{S\,I}+{\bf P}^{O\,I}=0$.

\color{red} \subsection{The reactive helicity flow } \color{black}
In turn, the reactive helicity flow [cf. Eq. (\ref{reactflow})] is
\be
\bm{\mathcal F}_{\qui} =\frac{c}{4k}[0, \frac{2Kq}{k^2}(| T_{ \parallel}|^2-|T_{\perp}|^2), -\frac{4q}{k}Re\{T_{\perp}T^{*}_{\parallel}\}]\exp(-2qz) \nonumber \\
= -\frac{1}{q}\,[0\,,\frac{2\pi K}{k^2} w_{react}\,,ck{\qui}].\,\,\,\,\,\,\,\,\,\,\,\,\,\,\label{spinreact}
\ee
 It should be emphasized that although both $Im\{{\bf S}\}$ and $\bm{\mathcal F}_{\qui} $ have an $y$-component proportional to $K>k$, (as well as proportional to $\qui$ in the former and to $w_{react}$ in the latter), there is no case of superluminal propagation for these quantities since both are alternating flows with zero time-average. We   show below that ${\cal F}_z$  represents {\it  up and down flow of reactive helicity} in the $OZ$ direction,  matching with the imaginary part of the complex helicity theorem, Eq. (\ref{bcpoy63}).

From the above equations it is important to remark  that while the time-averaged energy and helicity densities  are linked to $<{\bf S}>$, $<{\bf g}>$, $<\bm{\mathcal S}>$, $<{\bf P}^S>$ and $<{\bf P}^O>$,  {\it the densities of reactive power, $w_{react}$,  and reactive helicity, $\qui$, are exclusive of $Im\{{\bf S}\}$, ${\bf g}^{I}$ and $\bm{\mathcal  F}_{\qui}$}.

\color{red}
We   show below  that the $z$-component of $Im\{{\bf S}\}$, that matches with the imaginary part of the complex Poynting theorem, is associated with  an up and down flow of reactive power in the decay $z$-direction of the evanescent wave, and hence it is not a net flow of energy.  Analogously happens with the $z$-component of  $\bm{\mathcal F}_{\qui}$. However, as seen below,  both the $y$ and $z$-components of $Im\{{\bf S}\}$ produce  detectable optical forces and, hence, make the reactive quantities $ w_{react}$ and ${\qui}$ observable. \color{black}
  
\color{red} \subsection{Ractive power  conservation law } \color{black}

The $z$-component of $Im\{{\bf S}\}$, [cf. Eq. (\ref{ImPevan})],  depends on the {\it reactive  power} density, $\nabla\cdot Im\{{\bf S}\}$, of the evanescent wave in the half-space $z>0$,  (which actually concentrates in  the {\it near field} region above the interface $z=0$, namely at  $z<<\lambda$),  flowing back and forth along  $OZ$  at twice the frequency $\omega$,   without contributing to a net energy flow since its time-average is zero. I.e. one has from the CPV theorem:
 \be
 \nabla\cdot Im\{{\bf S}\} \equiv \partial_z Im\{{\bf S}\}=
 w_{react}  \nonumber \\
 \equiv -2q Im\{{\bf S}\}_z\,.\,\,\,\,\,\,\,\,\,\,\,\,\,\,\,\,\,\,\,\,\,\,\,\, \label{divPevan}
\ee
 Which obviously agrees with (\ref{ImPevan}). Therefore taking (\ref{divPevan}) into account, the total  {\it reactance}, \cite{balanis}  of the dielectric-air interface system associated to the evanescent wave is
\be
\int_V d^3 r  \nabla\cdot Im\{{\bf S}\}=\Sigma \int_{0}^{\infty}dz \,\nabla\cdot Im\{{\bf S}\}=-Im
\{{\bf S}_z(z=0)\}.\,\,\,\,\,\,\,\,\,\,\,\label{reactancia}
\ee
$\Sigma$ being the area of the $XY$-plane resulting from the volume integration.

\color{red} \subsection{Reactive helicity conservation law } \color{black}
Analogously,   the $z$-component of $\bm{\mathcal F}_{\qui}$, Eq. (\ref{spinreact}),  depends on the {\it reactive  helicity} density, $\nabla\cdot \bm{\mathcal F}_{\qui}$, of the evanescent wave in the half-space $z>0$,   concentrated in  the {\it near field} on  $z=0$, flowing up and down in the  the $z$-direction     without yielding  a net  flow since its time-average is zero. I.e. one has in agreement with Eq.(\ref{bcpoy53}), (${\bf j}=0$),
 \be
 \nabla\cdot \bm{\mathcal F}_{\qui} \equiv \partial_z \bm{\mathcal F}= \frac{2cq^2}{k^2} Re \{T_{\perp}T_{\parallel}^*\}\nonumber \\
 = -2q({\cal F}_{\qui})_z\equiv 2\omega {\qui}\,.\,\,\,\,\,\,\,\,\,\,\,\,\label{divReevan}
\ee

Hence,  the $z$-component of $\bm{\mathcal F}_{\qui}$ is proportional to the reactive power density. 

It is evident that while reactive power and reactive helicity exist in  evanescent waves, (see also Eqs.  (\ref{bcfor26b1}) of Section II.A and  (\ref{bcfora2021a}) of Section III.A), and in standing waves \cite{harrington,bliokh2}, they do not exist in plane propagating waves, whatever their polarization be. Therefore,  reactive helicity, like reactive energy, exists  in the near-field region of scattering or emitting objects.

\begin{figure*}[t!]
\begin{centering}
\includegraphics[width=18.5cm]{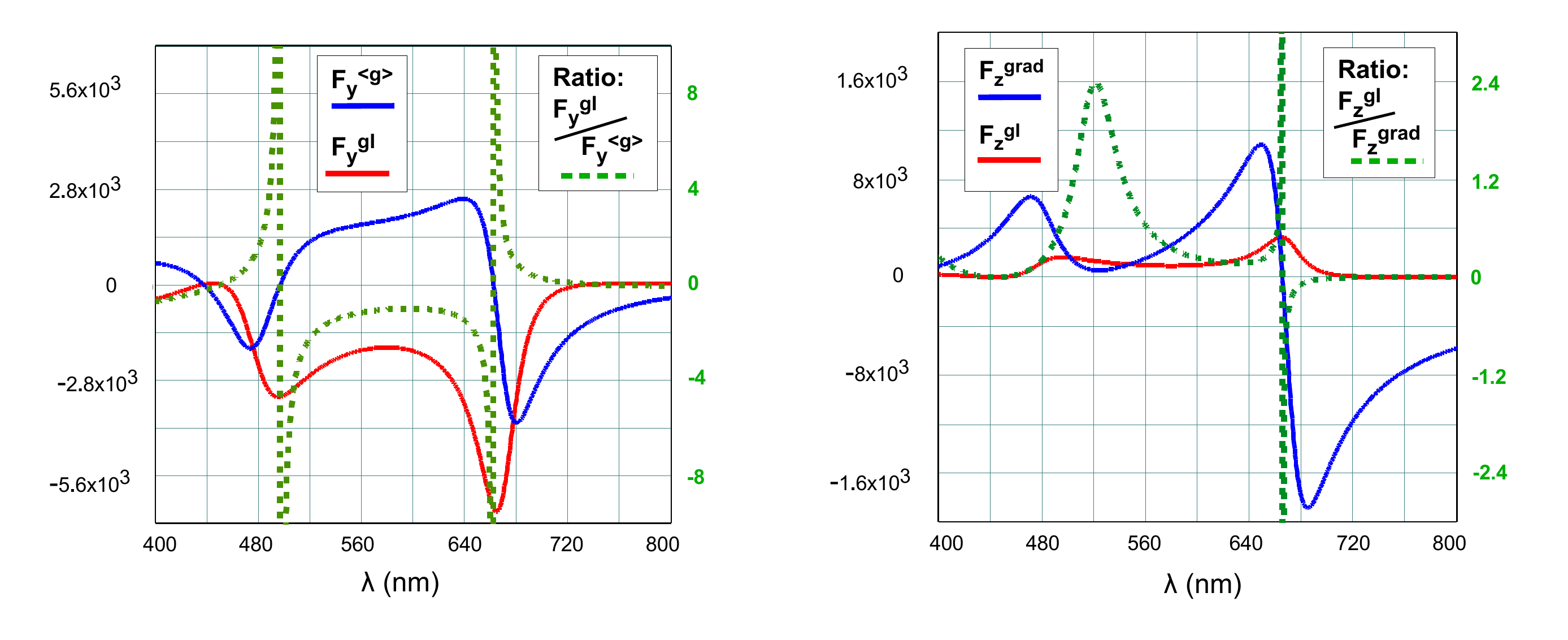}
\par\end{centering}
\caption{\color{red}  An evanescent wave, (see in Fig.1), is created in  $z\geq 0$  by TIR  at the interface $z=0$ of a linearly polarized plane propagating wave of complex amplitudes $A_{\perp}$ and $A_{\parallel}$ in the dielectric of refractive index $n=1.5$ with angle of incidence: $60^o$. A Si spherical  particle of radius $a=75$ nm is deposited on the interface. Its electric and magnetic dipole resonances are at $\lambda_e=492$ nm and $\lambda_m=668$ nm, respectively, [cf. Fig.4(left)]. This wave exerts on the particle the following transversal and perpendicular optical forces:
  (Left) With  $A_{\perp}=A_{\parallel}$, force transversal components,  $< {\bf F}_{e-m}^{{\bf g}^I }>_y$  due to ${ g}_y^I$, $< {\bf F}_{e-m}^{<{\bf g}> }>_y$  from $<{ g}_y>$, and ratio:   $< {\bf F}_{e-m}^{{\bf g}^I }>_y/< {\bf F}_{e-m}^{<{\bf g}> }>_y$ whose values (in green) are shown in the right  ordinate axis. (Right) Choosing $A_{\parallel}=0$, $A_{\perp}\neq 0$: Normal component $< {\bf F}_{e-m}^{{\bf g}^I }>_z$  due to ${ g}_z^I$,    gradient force $< {\bf F}^{grad} >_z$, and ratio:   $< {\bf F}_{e-m}^{{\bf g}^I }>_z/< {\bf F}^{grad} >_z$  whose values (in green) are shown in the right  ordinate axis. Notice that the positive values of the gradient force, repelling the particle from the interface, are due to negative values of $\alpha_e^R$ and/or  $\alpha_m^R$.  The bump  of $< {\bf F}^{grad} >_z$ near $\lambda_e$ and its  steep change of sign  close to $\lambda_m$, are due to the gradient force felt by the induced electric and magnetic dipoles, respectively.
The wavelength at which the second Kerker condition holds is $\lambda_{K2}=608$ nm. The force (normalized to area)  units are $fN/\mu m^2 $  for  an  incident power  of $1$ $mW/mm^2$.}
\end{figure*}

\color{red}\subsection{Observability of the transversal and perpendicular components of the imaginary momentum ${\bf g}^{I}$:  Optical forces on a magnetodielectric  dipolar particle  due to the reactive helicity and reactive power} 

 Let a magnetodielectric dipolar particle be placed on the $z=0$ interface. An illuminating wavefield, and in particular the evanescent wave, exerts an optical force on it  due to the interaction between its induced electric (e) and magnetic (m) dipoles \cite{nieto1}, viz.,
\be
< {\bf F}_{e-m} >= \frac{8\pi k^4 c}{3}[-Re(\alpha_e\alpha_m^*)<{\bf g}>+Im(\alpha_e\alpha_m^*){\bf g}^I].\,\,\,\,\,\,\,\,\,\ \label{Fem}
\ee 
 The e and m polarizabilities  $\alpha_e$ and $\alpha_m$ are related with the $a_1$ and $b_1$ Mie coefficients of the  field scattered by the particle by \cite{nieto1}:  $\alpha_e=i3a_1/2k^3$,  $\alpha_m=i3b_1/2k^3$.

The first term of (\ref{Fem}), proportional to $<{\bf g}>$, has been studied, (see its main features in Appendix C). Here we are interested in the second term that contains the reactive momentum ${\bf g}^I$. 

The  $y$-component $< {\bf F}_{e-m}^{{\bf g}^I }>_y$  due to ${ g}_y^I$  of (\ref{Fem}) was obtained in \cite{bliokh1}, being considered by the authors "a quite intriguing result" characterized  through  the second Stokes parameter with a rather small contribution in measurements of $< {\bf F}_{e-m} >_y$.    We have shown above that {\it this force  has a    reactive origin} since it  is fully due to the reactive spin $y$-component, {\it being characterized by the reactive helicity}  $\qui$. Furthermore, we establish here that $< {\bf F}_{e-m}^{{\bf g}^I }>_y$   is  detectable since  it may  widely exceed  the known component   $< {\bf F}_{e-m}^{<{\bf g}> }>_y$ due to $<g>_y$. 

For instance, Fig.2(Left) shows forces on a Si spherical particle placed on the interface $z=0$:   $< {\bf F}_{e-m}^{{\bf g}^I }>_y$    compared with $< {\bf F}_{e-m}^{<{\bf g}> }>_y$.  The particle electric and magnetic dipole resonances are at $\lambda_e=492$ nm and $\lambda_m=668$ nm, respectively, [see Fig.4(left)], and the {\it second Kerker condition (K2)} wavelength (at which the particle scatters minimum forward  intensity, see Section V.B) is $\lambda_{K2}=608$ nm. As seen,  the magnitude of  $< {\bf F}_{e-m}^{{\bf g}^I }>_y$ is much greter than $< {\bf F}_{e-m}^{<{\bf g}> }>_y$ near  $\lambda_e$ and $\lambda_m$ where the latter changes sign,  (see  the sharp  asymptotic values of the ratio between both forces). Thus $< {\bf F}_{e-m}^{{\bf g}^I }>_y$, which  keeps negative, should be detectable near these resonances.  Besides,  {\it in the proximities of K2, i.e. of $\lambda_{K2}$, $< {\bf F}_{e-m}^{{\bf g}^I }>_y=-< {\bf F}_{e-m}^{<{\bf g}> }>_y$. }

While  $< {\bf F}_{e-m}^{<{\bf g}> }>_z=0$, one observes in Fig.2(Right) features of  the perpendicular force $< {\bf F}_{e-m}^{{\bf g}^I }>_z$ similar to those of its $y$-component comparing it  with the gradient force  \cite{nietoOL},  which choosing $T_{\parallel}=0$ reads:  $< {\bf F}^{grad} >_z=-(1/2)q [\alpha_e^R|T_{\perp}|^2+\alpha_m^R|T_{\perp}|^2(2K^2/k^2-1)](0,0,1)$.  The superscript $R$ standing for real part. Again the sharp ratio between both near $\lambda=665$ nm indicates the wavelenght zone  where $< {\bf F}_{e-m}^{{\bf g}^I }>_z$, which remains positive, may be detected. Furthermore,  the bump of this ratio  in the proximities of $\lambda=520$ nm  shows that  this force is over twice the gradient force.

We conclude thereby that {\it there exists a measurable transverse  $y$-component of $< {\bf F}_{e-m} >$   due to  the reactive spin momentum density and hence to ${\qui}$, which may be dominant upon the  transversal component of $< {\bf F}_{e-m} >$  stemming from the field (Poynting) momentum, namely from ${\hel}$. Hence, ${\qui}$ is observable}. Besides, {\it the  normal force $< {\bf F}_{e-m} >_z $which is exclusively due ${g}_z^I$}, (since $<g_z>=0$), {\it  characterized by  the reactive power density $w_{react}$ of the evanescent wave has  not yet been addressed as far as we know, and may be detected at wavelengths at which, as seen above, clearly exceeds the gradient force, making $w_{react}$ also an observable quantity}
\color{black}

\section{Case 2: Reactive power and reactive helicity from a magnetodielectric dipolar sphere}
We  consider a magnetodielectric spherical particle of radius $a$ and volume $V_0$, dipolar in the wide sense, namely whose electric and magnetic polarizabilities are given by the first electric and magnetic Mie coefficients, respectively \cite{nieto1,nietoSi}, in air. We first address the reactive power and stored energy of  this magnetoelectric dipole with   electric and magnetic  moments ${\bf p}$ and ${\bf m}$, respectively.   For a  wave, ${\bf E}^{(i)}$,  ${\bf B}^{(i)}$, incident on the particle centered at ${\bf r}={\bf 0}$,
the dipolar moments are: ${\bf p}=\alpha_e{\bf E}^{(i)}{\bf (0)}$ and ${\bf m}=\alpha_m{\bf B}^{(i)}{\bf (0)}$. 

\subsection{Reactive power and  stored energy}
Concerning the CPV,  ${\bf S^{(s)}}=\frac{c}{8\pi}{\bf E}^{(s)}\times{\bf H}^{(s)\,*}$, of the emitted  fields, (cf. Appendix D),  we are interested in its radial component,  ${\bf S}^{(s)}\cdot \hat{\bf r}$,  across a spherical surface $\partial V$ of radius $r$  concentric with the particle and enclosing it. Using the fields of Appendix E, this is straightforwardly integrated, yielding 
\be
\int_{\partial V} {\bf S}^{(s)}\cdot \hat{\bf r}\,d^2 r=\frac{ck^4}{3}(|{\bf p}|^2+|{\bf m}|^2)+\frac{ick}{3r^3}(|{\bf p}|^2-|{\bf  m}|^2).\,\,\,\,\,\,\,\,\,\,\,\, \label{pow}
\ee
Whose real part, $\int_{\partial V} <{\bf S^{(s)}}>\cdot \hat{\bf r}\,d^2 r=\frac{ck^4}{3}(|{\bf p}|^2+|{\bf m}|^2) $ is  the well-known  radiated (scattered) total power, $W^{(s)}$, independent of  the distance $r$ to the center $r=0$, and  corresponds to the  wevefield re-radiated up to the far-zone, i.e.  that with the $r^{-1}$ dependence, [cf. Appendix D, Eqs. (D-1) and (D-2)].

\begin{figure*}[t!]
\begin{centering}
\includegraphics[width=18.0cm]{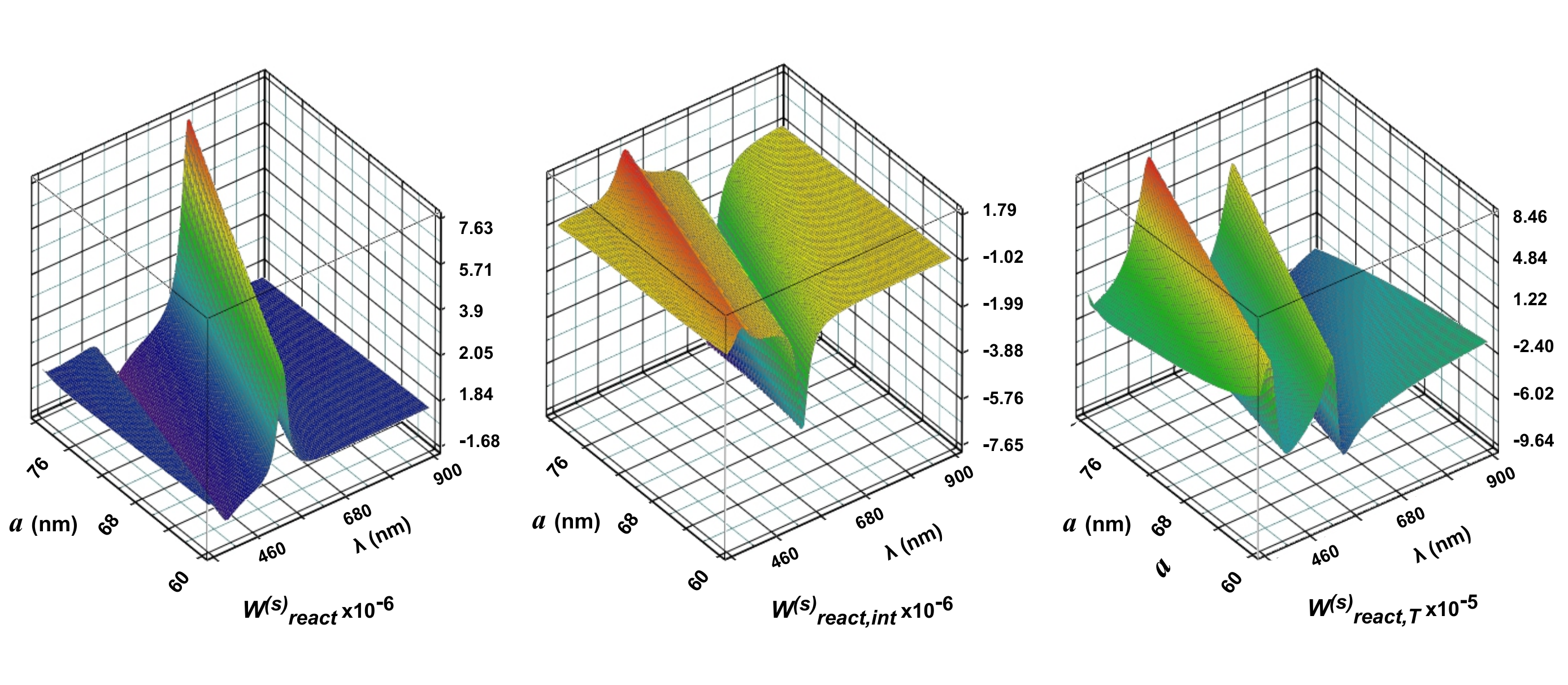}
\par\end{centering}
\caption{3-D graphs of reactive powers as functions of wavelength $\lambda$  and radius $a$ of a magnetodielelectric sphere  of Si in air,  illuminated by a (linearly or circularly polarized) plane wave.  (Left): External  $W_{react}^{(s)}$. (Center): Interior $W_{react,int}^{(s)}$. (Right): Total $W_{react,T}^{(s)}=W_{react}^{(s)}+W_{react,int}^{(s)}$. The  sign of  $W_{react,int}^{(s)}$ is opposite to that of  $W_{react}^{(s)}$ for all $\lambda$ and $a$. The dip and peak of    $W_{react}^{(s)}$ correspond to the electric and magnetic resonance, respectively, and are close to these resonances in 
$W_{react,int}^{(s)}$. Like  the resonances exhibited by the scattered power $W^{(s)}$ \cite{geffrin}, these peaks and dips are redshifted as $a$  increases. $W_{react,T}^{(s)}$ vanishes in the proximities of these resonances, irrespective of $a$.}
\end{figure*}

\begin{figure*}[t!]
\begin{centering}
\includegraphics[width=18.6cm,height=5.5cm]{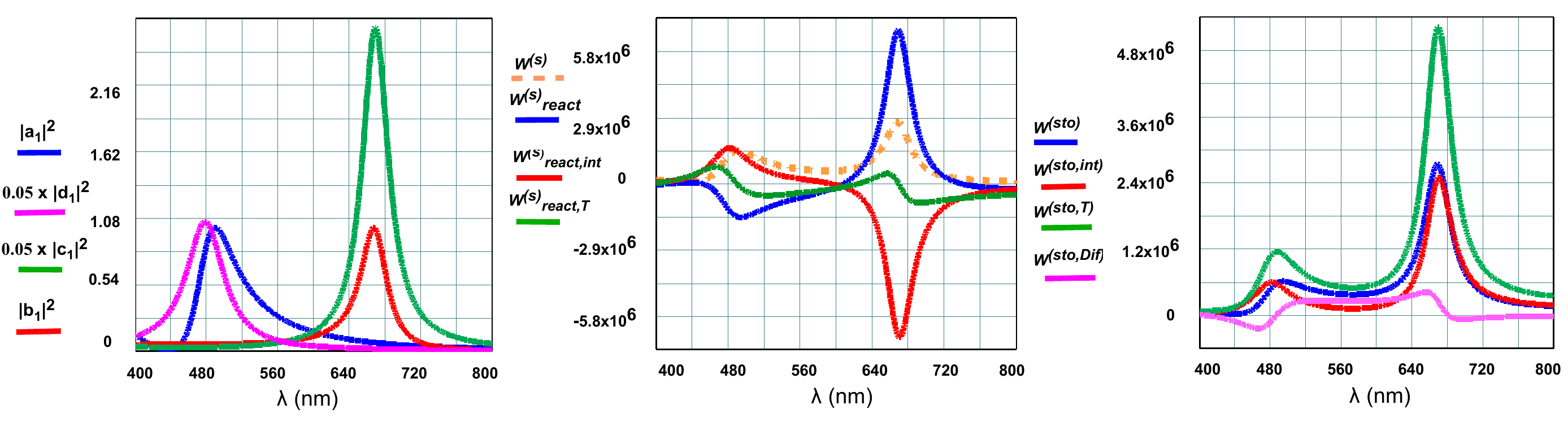}
\par\end{centering}
\caption{ (Left) Si sphere of radius $a=75nm$. (Left): Square moduli of the electric and magnetic Mie coefficients, $a_1$ and $b_1$, and $d_ 1$  and $c_1$ of the external and interior scattered  field, respectively.  (Center):  Scattered power $W^{(s)}$ and reactive  powers: external  $W_{react}^{(s)}$, interior $W_{react,int}^{(s)}$,  and  total $W_{react,T}^{(s)}=W_{react}^{(s)} +W_{react,int}^{(s)}$. Notice that $W_{react,T}^{(s)}$ vanishes at wavelenghts close to those of the electric and magnetic Mie resonances:  $\lambda_e=492 nm$ and  $\lambda_m= 668 nm$ where  $W^{(s)}$ is maximum . The lines $|a_1|^2$  and $|b_1|^2$ cross each other at the Kerker wavelengths: $\lambda_{K1}= 738.5 nm$, (at which  $W_{react}^{(s)}=0$ and  $W_{react, int}^{(s)}\simeq 0$),  and $\lambda_{K2}=608 nm$,  (where  both $W_{react}^{(s)}$ and  $W_{react, int}^{(s)}$ are near $0$). (Right): Stored energies: external $W^{(sto)}$, interior $W^{(sto,int)}$,  total $W^{(sto,T)}=W^{(sto)}+W^{(sto,int)}$, and difference $W^{(sto,Dif )}=W^{(sto)}-W^{(sto,int)}$..}
\end{figure*}
Here we are particularly interested in the imaginary part, 
\be
\int_{\partial V} Im\{{\bf S}^{(s)}\}
\cdot \hat{\bf r}\,d^2 r=\frac{ck}{3 r^3}(|{\bf p}|^2-|{\bf  m}|^2)]=-W_{react}^{(s)}. \,\,\,\,\,\,\,\,\,\,\,\,\label{repow}
\ee
Where $W_{react}^{(s)}$ is the {\it reactive power} outside  $V$. It arises from the near  and intermediate fields: ${\bf E}^{(s)}({\bf  r})=\{\frac{1}{\epsilon r^3}[3{\bf n}({\bf n}\cdot{\bf p})-{\bf p}] -\sqrt{\frac{\mu}{\epsilon}}({\bf n}\times{\bf m})
\frac{ik}{r^2}\}e^{ikr}$, and  ${\bf B}^{(s)}({\bf r})=\{\frac{\mu }{ r^3}[3{\bf n}({\bf n}\cdot{\bf m})-{\bf m}] +
\sqrt{\frac{\mu}{\epsilon}}({\bf n}\times{\bf p})\frac{ik}{r^2}\}e^{ikr}$.   (We recall that the expression: $\frac{ick}{3r^3}|{\bf p}|^2$ for a purely electric dipolar emitter  is well-known  in antenna theory \cite{harrington,stratton,balanis}). We note that the $r^{-3}$ dependence of $W_{react}^{(s)}$ makes it to acquire much larger values than $W^{(s)}$ in the near-field region.

The reactive power $W_{react}^{(s)}$ is the difference between averaged {\it stored} magnetic and  electric powers  which dominate in the near and intermediate-field regions around the particle, and that do not propagate. To see it, we write the mean stored electric and magnetic energy densities as \cite{collin,mcLean,geyi} $<\tilde{w}_e^{(s)}>=<{w}_e^{(s)}> -<{w}_e^{(FF)}>$,  $<\tilde{w}_m^{(s)}>= <{w}_m^{(s)}> -<{w}_m^{(FF)}>$. While  $<{w}_e^{(s)}>$ and  $<{w}_m^{(s)}>$ are the energy densities of the full fields ${\bf E}^{(s)}$ and ${\bf B}^{(s)}$, i.e with all terms of their expressions  (D-1)-(D-4) of Appendix D;   the supescript $(FF)$ stands for far-zone electric  and magnetic energy densities, i.e corresponding to the  fields (D-1)-(D-4) with only terms  of   $r^{-1}$ dependence. Since evidently  $<{w}_m^{(FF)}>=<{w}_e^{(FF)}>$, we have that $ <{w}_m^{(s)}> -<{w}_e^{(s)}>=<\tilde{w}_m^{(s)}>-<\tilde{w}_e^{(s)}>$  \cite{noteinterf}.

Then   we write   $W_{react}^{(s)}$ in terms of the  volume integral outside the volume $V$ in the above mentioned sphere $\partial V$ of radius $r$, centered in ${\bf r}={\bf 0}$:
\be
W_{react}^{(s)}=\frac{ck}{3 r^3}(|{\bf m}|^2-|{\bf  p}|^2)] \nonumber \\
=2\omega\int_{V_{\infty}-V}d^3 r \, (<{w}_m^{(s)}> -<{w}_e^{(s)}>)  \nonumber \\
=2\omega\int_{V_{\infty}-V}d^3r\, (<\tilde{w}_m^{(s)}>-<\tilde{w}_e^{(s)}>) . \label{Wwtot1}
\ee
 Where $V_{\infty}$ is the volume of a large sphere ($kr\rightarrow\infty$). Making $V=V_0$ and $r=a$, (\ref{Wwtot1})  yields the {\it  reactive power outside the particle}.
The proof of (\ref{Wwtot1}) is given in Appendix E, [cf. Eq. (\ref{stme})].  We anticipate that Eq.  (\ref{Wwtot1}) is a consequence of the {optical theorem for reactive power} which we put forward in  Section VI, [cf. Eq. (\ref{Icop4})].

As shown in Eq. (\ref{geninst}),   $\int_{\partial V} Im\{{\bf S}^{(s)}\}\cdot \hat{\bf r}\,d^2 r$ is associated to an instantaneous energy flow  alternating back and forth from the scatterer without losses, at frequency $2\omega$, with zero net energy transport in the  embedding vacuum.  Nonetheless, this alternating flow builds $W_{react}^{(s)}$ according to Eq. (\ref{repow}), also deduced from Eq. (\ref{Wwtot1}). As a consequence,  there is an accretion of   time-averaged  non-propagating {\it reactive power}  and {\it  stored  energy}, ${W}^{(sto)}$,   outside $V$. (We note  a concept analogous to  ${W}^{(sto)}$ for purely electric dipolar RF-antennas, cf.   e.g. \cite{harrington,collin,mcLean,geyi}). 
 Taking $V=V_0$, the {\it total  energy stored outside} the particle is  obtained by
\be
{W}^{(sto)}=\int_{V_{\infty}-V_0}(<\tilde{w}_m^{(s)}> +<\tilde{w}_e^{(s)}>) d^3 r\nonumber \\
=\int_{V_{\infty}-V_0}(<{w}_m^{(s)}> +<{w}_e^{(s)}>) d^3 r  \nonumber \\
-\int_{V_{\infty}-V_0}(<\tilde{w}_m^{(FF)}>+ <\tilde{w}_e^{(FF)}>) d^3 r \nonumber \\
=\frac{1}{6}(|{\bf p}|^2+|{\bf m}|^2) (\frac{1}{a^3}+\frac{2k^2}{a}).\,\,\,\,\,\, \,\,\,\,\,\,\,\
\label{Wtot1}
\ee
Appendix E provides the proof of  (\ref{Wtot1}), [cf. Eq.  (\ref{Wtot2})]. 

 The quality factor associated with  ${W}^{(sto)}$  is 
$
Q=\frac{2\omega{W}^{(sto)}}{W^{(s)}}=\frac{1}{(ka)^3}+\frac{2}{ka}. \, \label{factQ}
$
Hence being independent of the strength of the electric and/or magnetic dipole moments.

Since in general a dipolar particle in the wide sense cannot be abstracted as a point dipole,  the {\it overall interior reactive power},  $W_{react, int}^{(s)}$ and the {\it interior stored energy},  $W^{(sto, int)}$,  obtained analogously to (\ref{Wwtot1}) and (\ref{Wtot1}), but integrating  in $V_0$ the mean energies of the interior field, are also of interest. For a linearly, or circularly, polarized incident plane wave of unit intensity, a straightforward calculation yields  $(kc/3a^3)(9/4k^6)( 0.055|d_1|^2 \mp 0.018|c_1|^2)$. The upper and lower sign in $\mp$ apply to  $W_{react, int}^{(s)}$ and $W^{(sto, int)}$, respectively, while $d_1$ and $c_1$ are the first electric and magnetic Mie coefficients of the fields inside the  particle, (cf. e.g.  Eq.(4.45) of \cite{bohren}). Notice the appearance of $|d_1|^2$ and $|c_1|^2$  in $W_{react, int}^{(s)}$  with sign opposite to that of  $|{\bf p}|^2$ and $|\bf{m}|^2$ in  (\ref{Wwtot1}), i.e. of the first electric and magnetic  external Mie coefficient squared moduli,  $|a_1|^2$ and $|b_1|^2$,  in $W_{react}^{(s)}$; we shall see that this has consequences for the total (i.e. interior plus external) reactive power at resonant wavelengths.

\subsection{Reactive power, resonances, and Kerker conditions}
Fig.3  depicts $W_{react}^{(s)}(\lambda,a)$,  $W_{react, int}^{(s)}(\lambda,a)$,  and $W_{react,T}^{(s)}(\lambda,a)=W_{react}^{(s)}(\lambda,a)+W_{react,int}^{(s)}(\lambda,a)$ for Si spheres within the range of wavelengths where they become dipolar magnetodielectric, and for different size radii $a$, taking advantage of their scaling property with their impact parameter, (see Fig.2 of \cite{nietoSi}, and \cite{geffrin}). 
The redshift of the electric and magnetic dipole resonances $\lambda_e$ and $\lambda_m$ as $a$ grows, observed in the scattering cross-section \cite{geffrin,kivshar}, [see also  $W^{(s)}$ in  Fig.3(Center)], is observed in the peaks or dips of these reactive powers  as they are much influenced by these resonances. An important feature of these surfaces is that the total reactive power vanishes, changing its sign, close to the resonance wavelengths where $W^{(s)}$ is maximum, irrespective of $a$. This is detailed in Fig. 4(Center) on a cross sectional plane $a=75 nm$ of these surfaces, depicting the scattered power, along with the external, interior, and total reactive powers. Also Fig.4(Right)  shows the stored energies, while in Fig.4(Left)  one sees the electric and magnetic external  first Mie coefficients $a_1$ and $b_1$ with resonant  maxima at  $\lambda_e=492 nm$ and  $\lambda_m= 668 nm$,  respectively, and the internal coefficients,  $d_1$ and $c_1$.

{\it The vanishing of the total reactive power, $W_{react,T}^{(s)}$}, observed in Figs.3 (Right) and Fig.4(Center) {\it close to the electric and magnetic resonances $\lambda_e$ and $\lambda_m$ where the scattered power,  $W^{(s)}$, is produced with  maximum efficiency, is quite relevant because it manifests a cancellation between the internal and external reactive power peaks};   the  stored energies [cf. Fig. 4(Right)] being also resonant near these wavelengths.  In fact this matches  \cite{ziolkowski} with knowledge from RF antenna theory  according to which a maximum radiation efficiency is sought by minimizing their reactive power and $Q$-factor \cite{harrington,wheeler,chu,collin,mcLean,geyi,balanis}, even though in these works  the capacitive  - electric dipole wavelength $\lambda_e$:  ${\bf E}$ dominates in $r>a$, $r<\lambda_e$ - (inductive - magnetic dipole  wavelength $\lambda_m$:  ${\bf B}$ dominates in $r>a$, $r<\lambda_m$ -) nature of  $W_{react}^{(s)}$ is compensated and tuned to resonance by adding an inductive (capacitive) storage element.

Such element, here in the optics domain, is provided by the particle interior through the emergence  of a dominant ${\bf B}$ at $\lambda_e$ (dominant ${\bf E}$ at $\lambda_m$) of  $W_{react, int}^{(s)}$ inside the particle, ($r<a$).  By the same token the total stored energies have peaks in these  resonant wavelengths $\lambda_e$ and $\lambda_m$, like the scattered power $W^{(s)}$  [cf. Fig.4(Right)],  being evident  that the difference of external and internal stored energies:    $W^{(sto,Dif)}=W^{(sto)}-W^{(sto,int)}$ vanishes, like $W_{react,T}^{(s)}$ ,  close to  $\lambda_e$ and $\lambda_m$.

These results  illustrate  the concepts of reactive and stored power in and around a magnetodielectric, or Huygens, particle when choosing configurations and wavelengths such that the accretion of   external reactive power and stored energy  in the particle near-field,  through the IPV alternating flow,  be as large as possible; thus  scattering with maximum efficiency $W^{(s)}$, and possessing the highest possible $Q$-factor  for  applications in light-matter interactions, while its total reactive power $W_{react,T}^{(s)}$ vanishes or is near zero at resonant wavelengths.  Since, however, these  magnetodielectric nanoresonators  have rather low $Q$'s  ($Q\leq 7$), external (and internal) stored power enhancements may  be achieved  either by sets of such magnetodielectric particles, even metal coated, using them as building blocks of  photonic molecules, metasurfaces  \cite{kivsharmeta}, or in regimes  of  bound states in the continuum \cite{bonod,kivsharBIC}.

Returning to Fig. 4(Left), we observe the lines $|a_1|^2$ and  $|b_1|^2$ crossing each other at the two {\it Kerker wavelengths}: $\lambda_{K1}= 738.5 nm$ and $\lambda_{K2}=608 nm$, which correspond to the {\it first Kerker condition  (K1), (zero backscattering}, $\alpha_m=\alpha_e$, $|{\bf p}|=|{\bf m}|$), and {\it second Kerker condition (K2), (minimum forward scattering,}  $\alpha_m=-\alpha^*_e$,  $|{\bf p}|\simeq |{\bf m}|$),  \cite{kerker,nieto2011,nietoJNano,geffrin,lapin,staude,kivshar}. One sees in Fig. 4(Middle)  that at $\lambda_{K1}$,  $W_{react}^{(s)}=0$  in accordance with Eqs. (\ref{repow}) and (\ref{Wwtot1}),  $W_{react,int}^{(s)}\simeq 0$ ; while both  $W_{react}^{(s)}$ and $W_{react,int}^{(s)}\simeq 0$ at $\lambda_{K2}$. Also notice in Fig.4(Right) that the stored energies, $W^{(sto)}$ and $W^{(sto,int)}$, and scattered power, $W^{(s)}$, are near minimum in the proximities of  $\lambda_{K1}$ and $\lambda_{K2}$. These features happen in Fig.3 for any $a$.

Therefore, the analysis based on the particle reactive power allows  to envisage the Kerker conditions, K1 and K2, from a new standpoint: 

{\it The two Kerker conditions for a magnetodielectric dipolar particle  are those at which the total reactive power is near zero}. Namely, {\it the external reactive  power $W_{react}^{(s)}$ is  either zero (in K1), or   close to zero (in K2); while the internal reactive power is near zero both in K1 and K2}. In  consequence, {\it the  reactive  power  underlies the angular distribution of  scattered (or radiated) intensity and, hence,  the directivity of the magnetoelectric particle} in a  way complementary to that formerly addressed in Mie-tronics and  RF-antennas \cite{balanis, geyi}.

\begin{figure*}[t!]
\begin{centering}
\includegraphics[width=18.0cm]{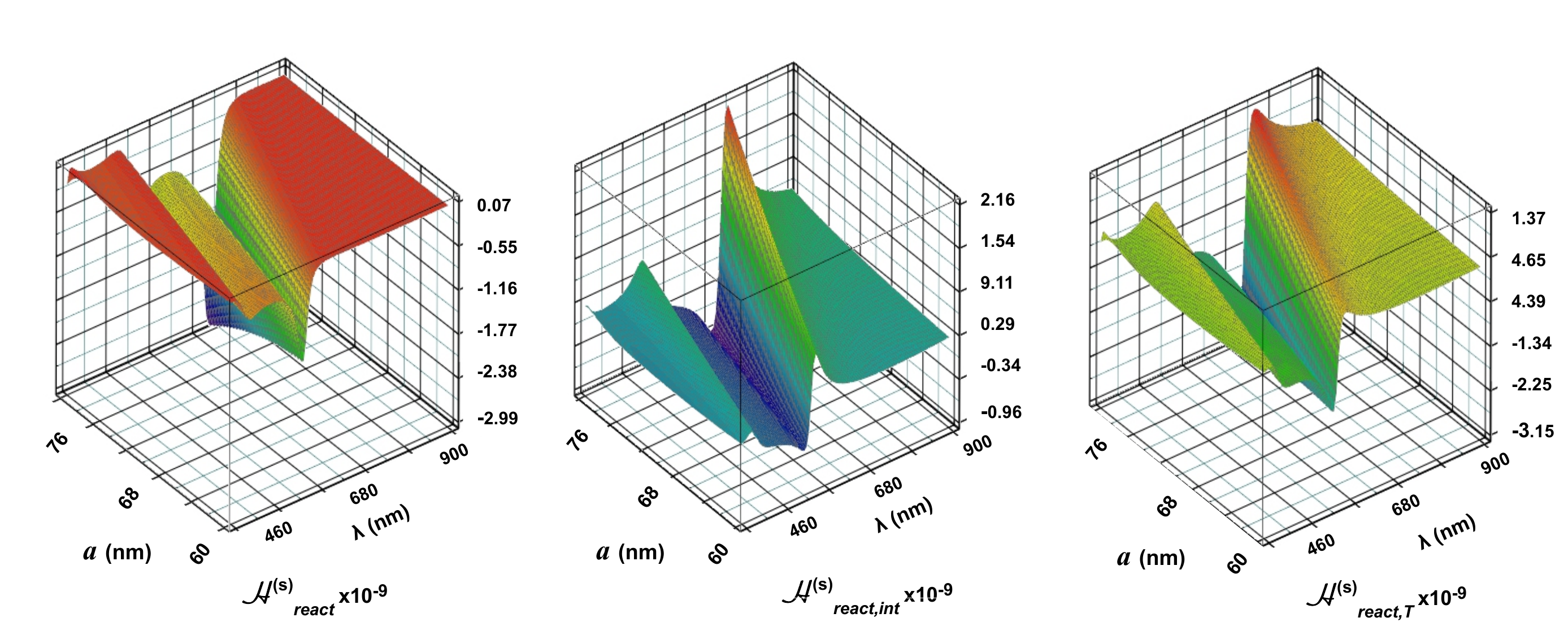}
\par\end{centering}
\caption{3-D graphs of reactive  helicities as functions of wavelength $\lambda$  and radius $a$ of a magnetodielelectric sphere  of Si in air,  illuminated by a left circularly polarized  incident plane wave, CPL(+).  (Left): External  ${\cal H}_{react}^{(s)}(\lambda,a)$. (Center): Interior ${\cal H}_{react,int}^{(s)}(\lambda,a)$. (Right): Total ${\cal H}_{react,T}^{(s)}(\lambda,a)={\cal H}_{react}^{(s)}(\lambda,a) +{\cal H}_{react,int}^{(s)}(\lambda,a)$. The two dips of   ${\cal H}_{react}^{(s)}$, and zero crossings of  ${\cal H}_{react,int}^{(s)}$,  are close to the electric and magnetic resonant wavelengths  $\lambda_e$ and $\lambda_m$,  and are redshifted with increasing $a$. Irrespective of the value of $a$,  ${\cal H}_{react}^{(s)}=0$ at $\lambda_{K1}$ and ${\cal H}_{react,T}^{(s)}$=0 close to $\lambda_e$ and $\lambda_m$.}
\end{figure*}
\begin{figure*}[t!]
\begin{centering}
\includegraphics[width=18.0cm]{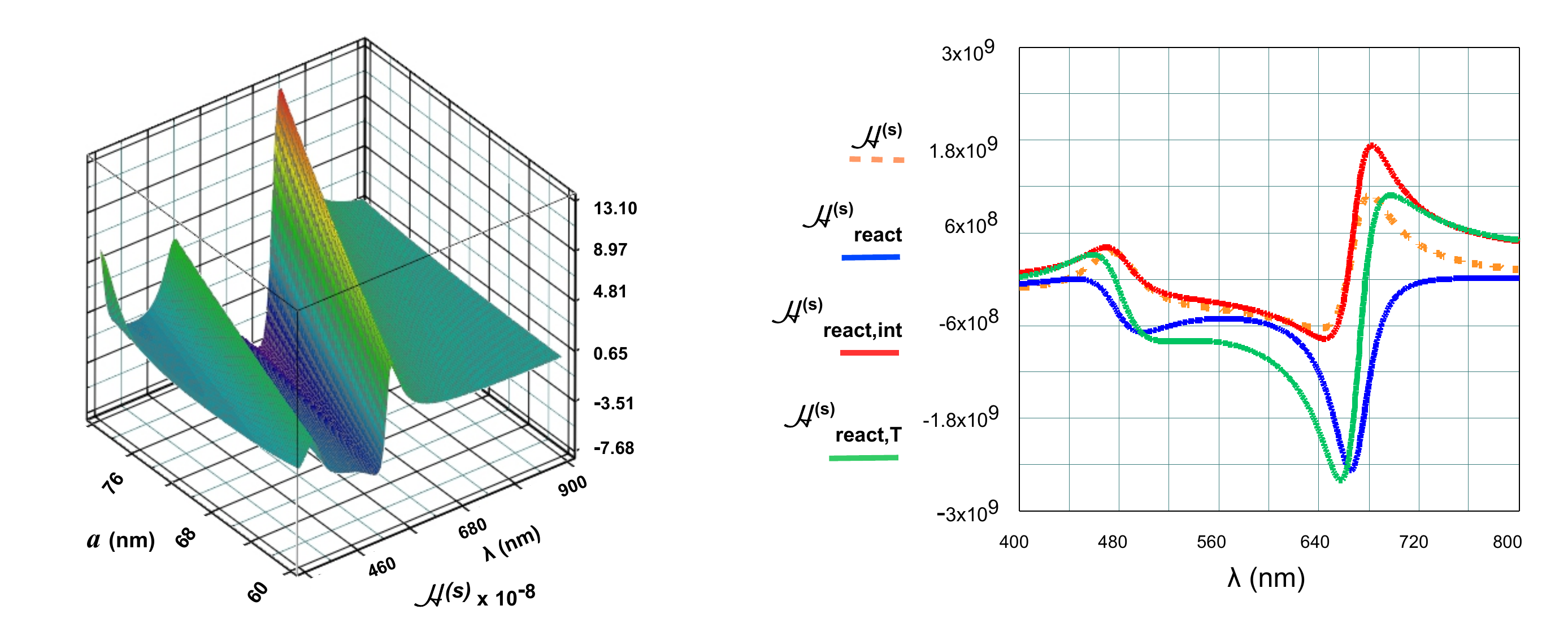}
\par\end{centering}
\caption{(Left) 3-D graph of the scattered helicity ${\cal H}^{(s)}(\lambda,a)$ generated by Si spheres in air,  illuminated by a left circularly polarized  incident plane wave, CPL(+).  It has two peaks,  at $475 nm$ and $679 nm$, influenced by the electric and magnetic resonant wavelengths  $\lambda_e=492 nm$ and $\lambda_m=668 nm$.  (Right): Making $a=75 
nm$,   ${\cal H}^{(s)}(\lambda)$, ${\cal H}_{react}^{(s)}(\lambda)$,  ${\cal H}_{react,int}^{(s)}(\lambda)$ and  ${\cal H}_{react,T}^{(s)}(\lambda)$.  At the Kerker wavelength $\lambda_{K1}=738.5 nm$, ${\cal H}_{react}^{(s)}$ vanishes. Also  ${\cal H}_{react,T}^{(s)}(\lambda)$ is zero  close to the wavelengths where the   scattered helicity  ${\cal H}^{(s)}$ is maximum, and in the vicinity of extrema of both  ${\cal H}_{react}^{(s)}$ and  ${\cal H}_{react,int}^{(s)}$.  }
\end{figure*}

\subsection{The reactive helicity }
Concerning the complex helicity flow of the fields scattered by the dipolar particle, [cf. Eq.(\ref{bcpoy63})],  $ \bm{\mathcal F}_C^{(s)} = \bm{\mathcal F}^{(s)}+i \bm{\mathcal F}_{{\qui}
}^{(s)}=(c/4k)Im\{({\bf B}^{(s) \,*} \times {\bf B}^{(s)}+{\bf E}^{(s)\,*} \times {\bf E}^{(s)})+i({\bf B}^{(s)\,*} 
\times {\bf B}^{(s)}-{\bf E}^{(s)\,*} \times {\bf E}^{(s)})\}$,  we obtain the flux 
$ \bm{\mathcal F}_C^{(s)} \,\cdot  \hat{\bf r}$ across the spherical surface $\partial V$ of radius $r\geq a$ centered in the particle. From Eqs. (\ref{app1}) and (\ref{app2}) for the fields in Appendix D, the terms that do not vanish on integration are
\be
\int_{\partial V} \bm{\mathcal F}_C^{(s)}\cdot \hat{\bf r}\,d^2 r=\frac{c}{4k}r^2\int_0^{2\pi}d\phi\int_0^{\pi}d\theta \, \hat{\bf r}\cdot  Im\{ [ \hat{\bf r}\times ({\bf m}^*  \times  \hat{\bf r})] \nonumber \\
\times ( \hat{\bf r} \times {\bf p})\frac{k^2}{r}(\frac{k^2}{r}+\frac{ik}{r^2})-C.C. \,\,\, \,\,\,\,\,\,\, \,\,\,\, \nonumber \\
+ [ \hat{\bf r}\times ({\bf p}^* \times  \hat{\bf r})] \times ( \hat{\bf r} \times {\bf m})\frac{k^2}{r}(\frac{k^2}{r}+\frac{ik}{r^2})-C.C.\}. \, \,\, \,\, \,\, \,\, \,\,\, \,\,\,\label{helflow1}
\ee
Where $C.C.$ denotes  complex-conjugated of the previous term. A straightforward calculation of (\ref{helflow1}) yields
\be
\int_{\partial V} \bm{\mathcal F}_C^{(s)}\cdot \hat{\bf r}\,d^2 r=\frac{8\pi}{3}ck^2 \{k\,Im[{\bf p}\cdot {\bf m}^*]+\frac{i}{r}Re[{\bf p}\cdot {\bf m}^*]\}, \,\, \,\, \,\,\,\, \,\, \,\,\, \,\,  \label{helflow2}
\ee
whose real part, ${\cal H}^{(s)}=\int_{\partial V} Re\{\bm{\mathcal F}_C^{(s)}\}\cdot \hat{\bf r}\,d^2 r =\int_{\partial V}\bm{\mathcal F} ^{(s)}\cdot \hat{\bf r}\,d^2 r=\frac{8\pi}{3}ck^3 Im[{\bf p}\cdot {\bf m}^*] $, is  the total helicity of the   scattered field, (as such, it coincides with Eq. (25) of \cite{nietoheli}). Like the scattered power, this helicity does not depend on the distance $r$.

However the imaginary part of (\ref{helflow2}), 
 \be 
-{\cal H}_{react}^{(s)}=\int_{\partial V} Im\{\bm{\mathcal F}_C^{(s)}\}\cdot \hat{\bf r}\,d^2 r =\int_{\partial V}\bm{\mathcal F}_{{\qui}}^{(s)}\cdot \hat{\bf r}\,d^2 r \nonumber \\
=\frac{8\pi}{3r}ck^2 Re[{\bf p}\cdot {\bf m}^*] ,\,\,\,\,\,\, \label{reraheli}
\ee
 comes from the interference of the intermediate field with $r^{-2}$ dependence and the far  field. ${\cal H}_{react}^{(s)}$ represents the external  {\it  reactive helicity},
\be
  {\cal H}_{react}^{(s)}=2\omega\int_{V_{\infty}-V} \qui^{(s)} \,d^3 r  \label{helrea1}
\ee
   outside $V$,  as detailed in its optical theorem discussed later, [cf.  Eq. (\ref{re_opteheli2})].  It decreases  as $r^{-1}$ as $r$ grows. Therefore,  in analogy with the reactive power,  in the near-field $(r<<\lambda$) the reactive helicity dominates upon  ${\cal H}^{(s)}$.   On making $V=V_0$, $r=a$, Eq.(\ref{helrea1}) becomes the overall reactive helicity outside the particle. 

It is of interest to specify the {\it overall reactive helicity of the field inside the particle},   ${\cal H}_{react,int}^{(s)}$, which is obtained integrating in $V_0$ the helicity of the internal field:  $2\omega\int_{V_{0}} \qui^{(s,int)} \,d^3 r$. For a  left circularly polarized incident plane wave of unit intensity, this quantity is equal to:   $(8\pi c k^2/3a)(9/4k^6) 0.234 \times 0.276 Re\{d_1  c^*_1\}$.

Figures 5  illustrate  ${\cal H}_{react}^{(s)}(\lambda,a)$,  ${\cal H}_{react,int}^{(s)}(\lambda,a)$, and  ${\cal H}_{react,T}^{(s)}(\lambda,a)={\cal H}_{react}^{(s)}(\lambda,a)+{\cal H}_{react,int}^{(s)}(\lambda,a)$, generated by Si spheres  illuminated by a left-circularly polarized, CPL(+), plane wave.  Again, the redshift with increasing radius $a$ is observed  in the electric and magnetic dipole resonant dips of  ${\cal H}_{react}^{(s)}$. An interesting feature of ${\cal H}_{react,int}^{(s)}$ is its similitude with ${\cal H}^{(s)}$, [cf. Fig.6(Center) and Fig.6(Left), as well as Fig.6(Right)], and specially their similarity with the converted helicity in the range between $\lambda_e$ and $\lambda_m$, (compare with Fig. 1(c) of  \cite{gutsche}). This remarks the contribution of the interior reactive helicity to the scattered helicity lineshape, and lays down an intriguing connection with previous studies \cite{poulikakos1,poulikakos2,gutsche} which attribute the conversion of helicity  to contributions of the scatterer volume and surface.

The total reactive helicity ${\cal H}_{react,T}^{(s)}$ plays on the  scattered helicity a role 
analogous to that of  the total reactive power on the scattered power. As seen in Fig. 6(Left)  ${\cal H}^{(s)}$ has resonant  dips influenced by the electric and magnetic resonant wavelengths, $\lambda_e$ and $\lambda_m$, of the scattered power. For $a=75$, [cf. Fig.6(Right)], these dips are at $\lambda=475 nm$ and $679nm$ and stem from the  resonances  at $\lambda_e=492 nm$ and $\lambda_m=668 nm$,  respectively.  On the other hand,  ${\cal H}_{react,T}^{(s)}$ becomes zero very close to these resonant wavelengths of ${\cal H}^{(s)}$.  We conclude, therefore, that {\it the interior reactive helicity ${\cal H}_{react, int}^{(s)}$ counteracts on the external reactive helicity ${\cal H}_{react}^{(s)}$ close the resonant wavelengths of ${\cal H}^{(s)}$, (where also   ${\cal H}_{react}^{(s)}$ and ${\cal H}_{react;int}^{(s)}$  are near  extreme values),  yielding a zero total reactive helicity ${\cal H}_{react,T}^{(s)}$}. This property may be observed in Fig.5(Right), independently of $a$, and it is clearly seen   in   Fig.6(Right).  We do not know, however, of any analogy of these effects in RF-antenna heory. 

\color{red} Notice  that these high index magnetoelectric particles have the interesting property of emitting a wavefield in which  there are not very large peaks of the total reactive power [cf. Fig.4(Center)]  versus those of the scattered (radiated) power; although, certainly, where this reactive power has extrema the scattered (radiated) power is well aside its peaks. However, under chiral ilumination these fields present high peaks of total reactive helicity versus its radiated one. For instance,   [see in Fig.6(Right)], the $75$ nm particle] yields a  the total reactive helicity with a large dip at 658$nm$ where the scattered (radiated) helicity  is near its minimum value. This occurs at shifted  positions as $a$ varies, [cf. Fig.6(Left) and Fig.5(Right)].

Therefore, the picture that emerges in these illustrations of such optical nanoantennas, considered as either primary or secondary sources,  is that the reactive power and the reactive helicity, which are concentrated both inside and in the near and intermediate regions of the source, have a hampering  effect in their   far-field scattering  (or radiation) efficiency. This is an analogous effect to that due to the presence of reactive power in RF-antennas.  In consequence, {\it if these nanoantennas emit chiral light, the total  reactive helicity hinders the efficiency of far-field scattered (or radiated) helicity, so that less of this helicity is emitted due to a build-up of reactive helicity  in and around the nanantenna}  \cite{bornseries}.\color{black}

 Under incident CPL  these particles are dual at $\lambda_{K1}$, and then   ${\bf p}_{\pm}=\pm i{\bf m}_{\pm}$   \cite{corbato,nietoheli}, the upper and lower sign applying to left circular, CPL(+) and right circular, CPL(-), respectively. Then ${\bf p}\cdot{\bf m}^*$ is purely imaginary at $\lambda_{K1}$ \cite{comment_CPL}, and one sees  from (\ref{helflow2}) that ${\cal H}_{react}^{(s)}=0$. This is observed in Fig.5(Left) for any $a$ and in Fig.6(Right) at $\lambda_{K1}=738.5 nm.$

 As $a$ varies,  ${\cal H}_{react}^{(s)}(\lambda,a)$   vanishes  at the corresponding Kerker wavelength $\lambda_{K1}$; this is seen in detail in Fig.6(Right) for $a=75nm$.

Therefore, {\it the first Kerker condition, K1, also  has the  novel property  that  under CPL illumination, the magnetodielectric particle}, which then becomes dual and hence emits a wavefield of well-defined helicity equal to the incident one, \cite{corbato,nietoheli,gutsche}, {\it does not generate external reactive helicity}.

\section{The reactive power optical theorem}

Consider a wavefield, ${\bf E}^{(i)}$, ${\bf H}^{(i)}$ incident on a  magnetodielectric  body of volume $V_0$. The field at any point of the embedding medium, (assumed to be vacuum or air), is represented as  ${\bf E}={\bf E}^{(i)}+{\bf E}^{(s)}$,  ${\bf H}={\bf H}^{(i)}+{\bf H}^{(s)}$, where the superscript $(s)$ denotes the scattered field.  Maxwell's equations  are written as:
\be
\nabla\times{\bf H}^{(i)}= -ik {\bf D}^{(i)}\,,\,\,\,\,\,\,\,\,\,\,\,\,\nabla\times{\bf E}^{(i)}= ik {\bf B}^{(i)};  \,\, \,\,\,\, \,\, \,\,\, \,\,\,\, \,\,\, \,\,\nonumber \\
\nabla\times{\bf H}^{(s)}= -ik {\bf E}^{(s)}-4\pi i k {\bf P}+\frac{4\pi}{c} {\bf j}\,,\,\,\,\,\,\,\,\,\,\,\,\nonumber \\
\nabla\times{\bf E}^{(s)}= ik {\bf H}^{(s)}+4\pi i k {\bf M}.  \,\, \,\,\,\, \,\, \, \,\, \,\,\,\, \,\, 
\label{maxwelleqs}
\ee
Where ${\bf P}$, ${\bf M}$ and ${\bf j}$ are the polarization, magnetization, and free current densities in $V_0$, respectively.  ${\bf D}^{(i)}={\bf E}^{(i)}$, ${\bf B}^{(i)}={\bf H}^{(i)}$.
We insert Eqs. (\ref{maxwelleqs}) into Eq.(\ref{bcpoy1}) for the total fields ${\bf E}$ and ${\bf H}$, and use the identity: $\nabla\cdot({\bf E}\times{\bf H}^*)= {\bf H}^* \cdot\nabla\times {\bf E}-{\bf E}\cdot \nabla\times{\bf H}^*$, integrating in a volume $V$ that contains  $V_0$, 
\be
\int_{V}d^3 r\, \{2i\omega(<w^{(i)}_m> -<w^{(i)}_e>)+i\frac{\omega}{4\pi} Re\{{\bf H}^{(i)}\cdot {\bf H}
^{(s)\,*} \nonumber \\
- {\bf E}^{(i)}\cdot {\bf E}^{(s\,)*} \}
+i\frac{\omega}{2}[ {\bf H}^{(i)\,*}\cdot {\bf M}-
 {\bf E}^{(i)}\cdot {\bf P}^{*}] \nonumber \\
-\frac{1}{2}\, {\bf j}^* \cdot{\bf E}^{(i)}+ \nabla \cdot {\bf S}^{(s)}\}d^3 r =-\frac{1}{2}\int_{V_0} d^3 r\, {\bf j}^* 
\cdot{\bf E}\,\,\, \,\,\,\,\,\, \nonumber \\ 
+2i\omega\int_{V}( <w^{}_m> -<w^{}_e>) d^3 r\,.\,\,\, \,\,\,\,\,\,\,\,\,\,\,\, \label{cop1}
\ee
If the incident field has no evanescent components, i.e. it is source-free and propagating, it does not store energy, so that the 
first term of (\ref{cop1}) is identically zero. However if it is evanescent , or it has evanescent components, we shaw in 
(\ref{divPevan}) that it stores reactive power and it is given by this term. Therefore we shall keep it in the above equation, which  is the {\it complex optical theorem}  in presence of the  scatterer.

If there are no  scattering induced sources other than ${\bf P}$ and ${\bf M}$ in the body, the free current ${\bf j}$ conveys the conversion of incident power into mechanical and/or thermal energy through the work done on the charges  $\frac{1}{2}Re\,\int_{V_0}d^3 r\,  {\bf j}^* \cdot ({\bf E} - {\bf E}^{(i)})$, which accounts for the decrease of energy ${\cal W}^{(a)}$ from the wave   as power absorbed by the body. Then taking the real part of  (\ref{cop1}) we obtain
\be
\frac{\omega}{2}\int_{V_0}d^3 r\,Im[ {\bf H}^{(i)\,*}\cdot {\bf M} +{\bf E}^{(i)\,*}\cdot {\bf P}] 
\nonumber \\ 
={\cal W}^{(a)}+ \int_{\partial V}d^2 r Re \{{\bf S}^{(s)}\}\cdot \hat{\bm r}. \label{Rcop1}
\ee
Which is the standard  {\it optical theorem} (OT) for energy  \cite{born, nieto1},  describing  the extinction of incident energy, [left side of (\ref{Rcop1})], and consequent absorption and radiation  of the total scattered energy. It reduces to its well-known expression \cite{nieto1} for dipolar particles on making ${\bf P}({\bf r})={\bf p}\,\delta({\bf r})$ and ${\bf M}({\bf r})={\bf m}\,\delta({\bf r})$, and then  $\int_{\partial V}d^2 r Re \{{\bf S}^{(s)}\}\cdot \hat{\bm r}=W^{(s)}$ in accordance with (\ref{pow}).

Here we are, however, interested in the imaginary part of   (\ref{cop1}),
\be
\int_{V}d^3 r\, [ 2\omega(<w^{(i)}_m> -<w^{(i)}_e>)+\frac{\omega}{4\pi} Re\{{\bf H}^{(i)}
\cdot {\bf H}
^{(s\,)*} \nonumber \\
- {\bf E}^{(i)}\cdot {\bf E}^{(s\,)*}\} +\frac{\omega}{2} Re \{ {\bf H}^{(i)\,*}\cdot {\bf M} -
 {\bf E}^{(i)\,*}\cdot {\bf P} \} ]   \nonumber \\
-\frac{1}{2}\,Im\{ \int_{V_0}d^3 r\,{\bf j}^* \cdot{\bf E}^{(i)}\}+\int_{\partial V} Im\{ {\bf S}^{(s)}\}\cdot \hat{\bm r} d^2 r\nonumber \\ 
 =-\frac{1}{2}\int_{V_0}d^3 r\, Im\{ {\bf j}^* \cdot{\bf E}\}\nonumber \\
+2\omega\int_{V}( <w^{}_m> -<w^{}_e>) d^3 r\,.\,\,\, \label{Icop1}\,\,\,\,\,\,\, \,\,\, \,\,\,\,\,\,
\ee
Which  becomes
\be
\frac{\omega}{2}\int_{V_0}d^3 r\,  Re \{ {\bf H}^{(i)\,*}\cdot {\bf M} - {\bf E}^{(i)\,*}\cdot {\bf P}\}  
=\nonumber \\ -\frac{1}{2}\int_{V_0}d^3 r\, Im\{ {\bf j}^*  \cdot{\bf E}^{(s)}\}
-\int_{\partial V} Im\{ {\bf S}^{(s)}\}\cdot \hat{\bm r} d^2 r
 +\nonumber \,\,\,\,\,\,\, \,\,\, \,\,\,\,\,\, \\
+2\omega\int_{V}( <\tilde{w}^{(s)}_m> -<\tilde{w}^{(s)}_e>) d^3 r\,.\, \label{Icop2}\,\,\,\,\,\,\, \,
\,\, \,\,\,\,\,\,\,\,\, \,\,\,\,\,
\,
\ee
 Where we have made use  of the fact that $<{w}_m^{(s)\,(FF)}>=<{w}_e^{(s)\,(FF)}>$ and hence $ <w^{(s)}_m> -<w^{(s)}_e>=<\tilde{w}_m^{(s)}>-<\tilde{w}_e^{(s)}>$, [cf. paragraph prior to Eq. (\ref{Wwtot1})]. 

Equation (\ref{Icop2}) is our formulation of the {\em reactive power optical theorem}  (ROT) for a generic scatterer whose response to illumination induces densities of polarization ${\bf P}$, magnetization ${\bf M}$, and  free current ${\bf j}$,  \cite{comment ROT}.  The left side constitutes the extinction of incident energy which produces the build-up of  external  reactive power on scattering in the right side of (\ref{Icop2}). Thus Eq. (\ref{Icop2}) describes how, in addition to being  radiated into the far zone as  Eq. (\ref{Rcop1}) illustrates, scattering gives rise to non-radiated energy, stored in  $V$ in the form of (external) reactive power, flowing out from the scattering object  and returning to it.   

For a dipolar particle the ROT reduces to
\be 
\frac{\omega}{2} Re \{ {\bf H}^{(i)\,*}(0)\cdot {\bf m} - {\bf E}^{(i)\,*}(0)\cdot {\bf p}\}   
=\,\,\,\,\,\,\,\,\,\,\nonumber \\-\frac{1}{2}\int_{V_0}d^3 r\, Im\{ {\bf j}^*  \cdot{\bf E}^{(s)}\}
-\int_{\partial V} Im\{ {\bf S}^{(s)}\}\cdot \hat{\bm r} d^2 r\,\,\,\,\,\,\,\,\,\,\nonumber \\
+2\omega\int_{V}( <\tilde{w}^{(s)}_m> -<\tilde{w}^{(s)}_e>) d^3 r\,.\,\,\,\,\,\,\,\,\,\,\,\,\, \label{Icop21}
\ee
The argument $0$ indicates that the fields are evaluated at the particle center $r=0$.

Concerning a dipolar particle, the process described by the complex optical theorem (\ref{cop1}) is analogous  to  that in which the feeding energy from an alternate current $I$, induces  an  oscillating  dipole in a small antenna,  which emits radiated and stored power  through the extinction, $\frac{1}{2} Z I^2$, of the driving energy. $Z$ being the antenna input impedance, $Z=R_l+R_r-iX$.  The dipole loss  resistance $R_l$ and {\it radiation resistance} $R_r$ \cite{balanis}, generated in accordance with the optical theorem  (\ref{Rcop1}), are  $2W^{(a)}/|I|^2$ and   $2W^{(s)}/|I|^2$, respectively. On the other hand, the dipole {\it reactance} $X$ (which for a magnetoelectic dipole is either capacitive or inductive \cite{jackson}, depending on the wavelength $\lambda_e$ or $\lambda_m$) stems from its external  reactive power, $W_{react}^{(s)}$, \cite{balanis} whose generation is ruled by the ROT (\ref{Icop21}). Hence, in this  context the extinction term in the left sides of (\ref{Rcop1}) and (\ref{Icop21}) may be associated to $\frac{1}{2}( R_l+R_r) |I|^2$ and $-\frac{1}{2} X |I|^2$, respectively.

On taking  $\partial V$ and $V$ as $\partial V_{\infty}$ and $V_{\infty}$,  respectively, corresponding to a large sphere,   ($kr\rightarrow \infty$), the flux of   scattered CPV   
across $\partial V_{\infty}$ is real and equals the total scattered energy $W^{(s)}$. Therefore, Eq.  (\ref{Icop21}) yields
\be
\frac{\omega}{2} Re \{ {\bf H}^{(i)\,*}(0)\cdot {\bf m} - {\bf E}^{(i)\,*}(0)\cdot {\bf p}\}=
 -\frac{1}{2}\int_{V_0}d^3 r\, Im\{ {\bf j}^*  \cdot{\bf E}^{(s)}\}\nonumber  \\
+2\omega\int_{V_{\infty}}( <\tilde{w}^{(s)}_m> -<\tilde{w}^{(s)}_e>) d^3 r\,.\,\,\,\,\,\,\,\,\,\,\,\, \label{Icop3}
\ee
Which introduced into  (\ref{Icop21}) leads to
\be
-\int_{\partial V} Im\{ {\bf S}^{(s)}\}\cdot \hat{\bm r} d^2 r =\,\,\,\,\,\,\,\,\, \,\,\,\,\,\,\nonumber \\
2\omega\int_{V_{\infty}-V}( <\tilde{w}^{(s)}_m> -<\tilde{w}^{(s)}_e>) d^3 r=W_{react}^{(s)} \,.\, 
\label{Icop4} 
\ee
In contrast with its real part,  the flow (\ref{Icop4}) depends on the integration domains, $V$ and $\partial V$. Notice that  (\ref{Icop4}) conincides with Eq. (\ref{Wwtot1}).

\section{The reactive helicity optical theorem}
Next, we put forward the law which rules the formation of external reactive  helicity by scattering   in the near and intermediate-field regions of the particle through extinction of helicity of the incident wave. Let us  introduce  Eqs. (\ref{maxwelleqs}) into the identities: $\nabla\cdot({\bf H}^*\times{\bf H})= {\bf H} \cdot\nabla\times {\bf H}^*-{\bf H}^*\cdot \nabla\times{\bf H}$ and  $\nabla\cdot({\bf E}^*\times{\bf E})= {\bf E} \cdot\nabla\times {\bf E}^*-{\bf E}^*\cdot \nabla\times{\bf E}$, integrating in a volume $V$ that contains the scattering volume $V_0$. With  ${\bf E}={\bf E}^{(i)}+{\bf E}^{(s)}$,  ${\bf H}={\bf H}^{(i)}+{\bf H}^{(s)}$, adding the respective expressions, employing the conservation equation (\ref{bcpoyReal}) and using the definitions  (\ref{heli}) and  (\ref{helflow}), ($n$=1 and ${\bf H}={\bf B}$ outside $V_0$), we arrive at
\be
2\pi c\int_{V_0}d^3r Re \{{\bf E}^{(i)\,*}\cdot{\bf M}-{\bf H}^{(i)\,*}\cdot{\bf P}\}= \nonumber \\
\int_{\partial V}d^2 r \bm{\mathcal F}^{(s)}\cdot {\bf n} +\frac{2\pi}{k}\int_{V_0}d^3r Im\{{\bf H}^{(s)\,*}\cdot{\bf j}\}, \label{opteheli}
\ee 
which is the known {\it optical theorem for the electromagnetic helicity} \cite{nietoheli} applying to  magnetodielectric arbitrary scattering bodies. 
\color{red}
Notice that for a dipolar particle, since   ${\bf P}({\bf r})={\bf p}\,\delta({\bf r})$, ${\bf M}({\bf r})={\bf m}\,\delta({\bf r})$ and $\int_{\partial V}d^2 r \bm{\mathcal F}^{(s)}\cdot {\bf n}=(8\pi c k^3/3) Im [{\bf p}\cdot{\bf m}^*]$, Eq. (\ref{opteheli}) becomes like Eq. (28) of \cite{nietoheli}.

\color{black}
However our focus is the conservation law of the reactive helicity. To formulate it in the form of an optical theorem we substract, rather than add, the above vector identities, and make use of the conservation law (\ref{bcpoy53}) along with definitions (\ref{reactheli}) and (\ref{reactflow}). Then, proceeding as before, it is straightforward to obtain 
\be
2\pi c\int_{V_0}d^3r Re \{{\bf E}^{(i)\,*}\cdot{\bf M}+{\bf H}^{(i)\,*}\cdot{\bf P}\}= \nonumber \\
 -\frac{2\pi}{k}\int_{V_0}d^3r Im\{{\bf H}^{(s)\,*}\cdot{\bf j}\}-\int_{\partial V}d^2 r \bm{\mathcal F}_{{\qui}}^{(s)}\cdot {\bf n}
\nonumber \\
+2\omega \int_{V}d^3r \qui^{(s)}, \,\,\,\,\,\,\,\,\,\,\,\,\,\, \,\,\,\,\,\label{re_opteheli}
\ee
Equation (\ref{re_opteheli}) is the {\it reactive helicity optical theorem} and applies to a  generic magnetodielectric scatterer \cite{re opteheli}.  The left side represents the extinction of helicity of the incident wave on build-up outside the body  of a reactive helicity by scattering, given by the right side of (\ref{re_opteheli}). Thus, like the energy, the incident helicity gives rise to a reactive one associated to the scattered field, which, in addition to the internal reactive helicity, is  stored around the particle, dominating in the near and intermediate-field regions where it flows back and forth from the scatterer. 

This storage is  seen  by first considering $V$ to be $V_{\infty}$ in (\ref{re_opteheli}). Then  [cf. \cite{comment Vinfty}]
\be
\int_{\partial V_{\infty}}d^2 r \bm{\mathcal F}_{{\qui}}^{(s)}\cdot {\bf n}=0.
\ee
Therefore,
\be
2\pi c\int_{V_0}d^3r Re \{{\bf E}^{(i)\,*}\cdot{\bf M}+{\bf H}^{(i)\,*}\cdot{\bf P}\}= \nonumber \\
 -\frac{2\pi}{k}\int_{V_0}d^3r Im\{{\bf H}^{(s)\,*}\cdot{\bf j}\}
+2\omega \int_{V_\infty}d^3r \qui^{(s)}, \,\,\,\,\,\,\,\,\,\,\,\,\,\, \,\,\,\,\,\label{re_opteheliinfty}
\ee
which substituted in (\ref{re_opteheli}) leads to
\be
-\int_{\partial V}d^2 r \bm{\mathcal F}_{{\qui}}^{(s)}\cdot {\bf n}=
2\omega \int_{V_{\infty-V}}d^3r \qui^{(s)}. \,\,\,\,\,\,\,\,\,\,\,\,\,\, \,\,\,\,\,\label{re_opteheli2}
\ee
Notice that if $V=V_0$, Eq. (\ref{re_opteheli2}) accounts for the reactive helicity stored outside the scattering body. 

Likewise, if the scatterer is a dipolar particle, the extinction term in the left side of (\ref{re_opteheliinfty}) becomes $2\pi c Re \{ {\bf E}^{(i)\,*}(0)\cdot {\bf m} + {\bf H}^{(i)\,*}(0)\cdot {\bf p}\}$. Then (\ref{re_opteheli2}) coincides with ${\cal H}_{react}^{(s)}$, Eq. (\ref{helrea1}) according to (\ref{reraheli}), thus proving it; and   illustrates how the  external reactive helicity is stored around the particle without being scattered into the far-zone.

\section{Consequence of the reactive power optical theorem: Significance of the reactive helicity in reactive dichroism}

In dichroism,  chiral light  illuminates  a   chiral particle,  molecule, or nanostructure. Assuming it dipolar, its  constitutive relations for the induced dipole moments, $ {\bf p}$ and  ${\bf m}$, and the  incident field are
\be
{\bf p}=\alpha_{e} {\bf E}^{(i)}+\alpha_{em}{\bf  B}^{(i)}, \,\,\,\,\,
{\bf m}=\alpha_{me}{\bf E}^{(i)}+\alpha_{m}{\bf B}^{(i)}.\,\,\,\,\,\,\,\,\,\, \label{consti}
\ee
The  electric,  magnetic, and magnetoelectric polarizabilities being  $\alpha_{e}$,  $\alpha_{m}$,  $\alpha_{em}$,  and $\alpha_{me}$; and fulfilling   $\alpha_{em}=-\alpha_{me}$ since the object is chiral \cite{tang,barron}.

The  signal  re-emitted (or scattered) by the excitation of this dipolar body  discriminates enantiomers (i.e. particles with either $\alpha_{me}$  or  $-\alpha_{me}$) \cite{tang,barron,schellman} by using as rate of excitation:
${\cal W}^{(s)}=\frac{\omega}{2} Im\{{\bf E}^{(i)\,*}(0)\cdot{\bf p}+{\bf B}^{(i)\,*}(0)\cdot{\bf m}\}$; [cf. left side of  Eq. (\ref{Rcop1})].

For instance, consider the pair of illuminating fields  \cite{tang}: ${\bf \cal E}^{(i)}({\bf r},t)=Re [\pm{\bf E}^{(i)}({\bf r}) \exp(-i\omega t)]$ and  ${\bf \cal H}^{(i)}({\bf r},t)=Re [{\bf H}^{(i)}({\bf r}) \exp(-i \omega t)]$, whose respective helicites are: ${\hel}_i^{+}$ and  ${\hel}_i^{-}$, with ${\hel}_i^{+}={\hel}_i= -{\hel}_i^{-}$. On employing (\ref{Rcop1}) and (\ref{consti}), the particle excitation rate  becomes: ${\cal W}^{(s)\,\pm}=\frac{\omega}{2}\{\alpha_e^{I} |{\bf E}^{(i)}(0)|^2+\alpha_m ^{I} |{\bf B}^{(i)}(0)|^2 \pm  2 \alpha_{me}^{R}\,Im [{\bf E}^{(i)}\cdot{\bf B}^{(i)*\,}]\}$$=\frac{\omega}{2}\{\alpha_e^{I} |{\bf E}^{(i)}(0)|^2+\alpha_m ^{I} |{\bf B}^{(i)}(0)|^2 \pm  4k \alpha_{me}^{R}\,\hel_i\}.$ 
 The superscripts  $I$ and $R$   denote imaginary and real part, respectively. 
Clearly, the sign $+$ or $-$ appears according to whether the helicity of the illumination is positive: ${\hel}_i^{+}={\hel}_i$, or negative: ${\hel}_i^{-}=-{\hel}_i$. 

Then the above expression   of ${\cal W}^{(s)\,\pm}$ yields the well-known dissymmetry factor $g=({\cal W}^{(s)
\,+}-{\cal W}^{(s)\,-})/({\cal W}^{(s)\,+}+{\cal W}^{(s)\,-})$   \cite{barron,schellman} proportional to 
$ \alpha_{me}^{R}\,\hel_i$ \cite{tang}.

However, the ROT establishes that  rather than the power radiated in the far-zone  (\ref{Rcop1}), {\it one may address the excitation of  stored reactive power, which dominates in the  near and intermediate-field  regions of the particle}, which is given by the left side of Eq. (\ref{Icop21}), viz. $ {\cal W}_{react}^{(s)}=\frac{\omega}{6} \Re\{{\bf E}^{(i)\,*}(0)\cdot{\bf p}-{\bf B}^{(i)\,*}(0)\cdot{\bf m}\}$. Then using  the above pair of illumination fields and Eq. (\ref{consti}), one obtains the discriminatory reactive power
\be
{\cal W}_{react}^{(s)\,\pm}= \frac{\omega}{2}\Re\{\pm{\bf E}^{(i)\,*}(0)\cdot{\bf p}-{\bf B}^{(i)\,*}(0)\cdot{\bf m}\} \,\,\,\,\,\,\,\,\,\,\,\,\,\,\,\,\,\,\,\,\,\,\,\,\,\,\,\,\,\,
\nonumber \\ 
= \frac{\omega}{2}\{\alpha_e^{R} |{\bf E}^{(i)}|^2-\alpha_m ^{R} |{\bf B}^{(i)}|^2 \mp  2 \alpha_{me}^{R}\,\Re [{\bf E}^{(i)}\cdot{\bf B}^{(i)*\,}]\}\ \,\,\,\,\,\,\,\,\,\,\, \nonumber \\
 =\frac{\omega}{2}\{\alpha_e^{R} |{\bf E}^{(i)}|^2-\alpha_m ^{R} |{\bf B}^{(i)}|^2 \mp  4k \alpha_{me}^{R}\,\qui_i\}.\,\,\,\,\,\,\,\,\,\,\,     \label{reacdicroi1}
\ee
So that now the  sign  $-$ or $+$ applies according to whether the reactive helicity of the illumination is positive: ${\qui}_i^{+}={\qui}_i$, or negative: ${\qui}_i^{-}=-{\qui}_i$, respectively; and (\ref{reacdicroi1}) yields a dissymmetry factor proportional to   $-\alpha_{me}^{R}\,\qui_i $ \,\cite{comment2}.

We propose  Eq. (\ref {reacdicroi1}) as the basis of {\it reactive dichroism} observations. At difference with standard dichroism, it involves a chiral incident field with non-zero reactive helicity, and constitutes a {\it near-field optics} technique.

Therefore, \color{red}  in an analogous way as detecting the radiated energy ${\cal W}^{(s)\,\pm}$, (or absorption/extinction energy), in standard dichroism  involves the helicity $\hel_i$ of the incident wave, {\it in  reactive dichroism, observing the excitation of reactive power ${\cal W}_{react}^{(s)\,\pm}$ in the chiral particle conveys the  incident  field reactive helicity  $ \qui_i$} ,  which evidently comes out on using (\ref{reacdicroi1})  in  a {\it dissymmetry factor} defined as $g_{react}=2( {\cal W}_{react}^{(s)\, +}-{\cal W}_{react}^{(s)\, -})/( {\cal W}_{react}^{(s)\, +}+{\cal W}_{react}^{(s)\, -})$. Notice that  being $ \qui_i$ measurable, so is   $g_{react}$ in proportion to $ \qui_i$.
\color{black}

There is a variety of pairs of illumination wavefields  that,  like  in the  above  illustration,  are interchangeable
by parity, and that one may employ in experiments. As discussed in previous sections, non-propagating fields in free-space  possess a non-zero $\qui_i$; e.g. elliptically (or circularly,  in particular) polarized standing waves, near fields from an emitter, or evanescent and other surface waves,  fulfill Eq. (\ref{reacdicroi1}).

\section{Conclusions}

Given the broad interest of evanescent waves at the nanoscale, and of small particles as light emitting \color{red}nanoantennas, \color{black} couplers, and metasurface elements, the contributions of this paper on its reactive quantities  is summarized in the following main conclusions:

(1) We have established the concepts of  {\it complex helicity} density and its   {\it complex helicity flow}, together with their conservation law that we name  {\it complex helicity theorem}.  Its real part is the well-known   conservation equation  of optical helicity, while its imaginary is a novel  law that governs the build-up of   {\it reactive helicity} through  its imaginary, and thus zero time-average,  flow.  The concept of reactive helicity density unifies  that of magnetoelectric energy density,  previously introduced  by  symmetry arguments, and the so-called  real helicity. In this way, we put forward its  {\it conservation law and observability, thus completing  the fundamentals of this quantity}. 

(2) The conservation of reactive helicity and  reactive  power, and their zero time-averaged flow,  has been illustrated in two paradigmatic systems: an evanescent wave and a wavefield scattered (or emitted) by a dipolar magnetodielectric particle.  For the former we have put forward   \color{red}{\it  reactive orbital and spin momenta} that characterize its imaginary (reactive) field (Poynting) momentum; showing that {\it the wave density of reactive helicity is observable from an experimentally detectectable mixed electric-magnetic transversal optical force } exerted by the  evanescent wave on a small high refractive index particle (which behaves as magnetodielectric)   {\it  through this  reactive Poynting momentum}. On the other hand, {\it we have uncovered a novel non-conservative force in the decay direction of the evanescent wave, which can be discriminated from the gradient one and thus detected, making observable the wavefield reactive power density.}
 \color{black} 

(3) Concerning the stored energy, reactive power, and reactive helicity of the field scattered by a dipolar magnetodielectric particle,   we have shown that {\it they provide a novel framework to study  the particle emission directivity}. This has been illustrated on  addressing the   two Kerker conditions, K1 of zero backscattering, and K2 of  minimum forward scattering. We have established that under CPL  incident light,  {\it the external reactive  power  is  zero in K1 or   close to zero in K2; while the internal reactive power, and hence the total reactive power,  is near zero both in K1 and K2}. Also,  we have proven an additional novel property of the particle at K1 wavelengths, namely, it produces a scattered field {\it with nule overall external reactive helicity}.

(4) We have established a  {\it  reactive helicity optical theorem} that  governs the {\it build-up and storage of near-field reactive helicity}, on extinction of the incident helicity as light interacts with a generally magnetodielectric  \color{red} nanoantenna.  \color{black}Also we have shown that {\it the emission of  resonant  scattered power and of resonant scattered helicity coincides with a nule, or near zero,  total reactive power and  helicity, respectively}, i.e. those given by the sum of the overall interior and external reactive powers and helicities. \color{red}Conversely, peaks of total reactive helicity (reactive power) are associated with poorer efficiency in the emission of radiated (or scattered) helicity (power). \color{black}

(5) A  {\it  reactive power optical theorem} has been put forward. It  rules the {\it formation of external reactive power} and  {\it  stored electric and magnetic energies}, which dominate in the near and intermediate-field zones of a   magnetodielectric particle by extinction of the illuminating energy.  

(6) This latter theorem  provides a framework to studying the {\it near-field response of  chiral  nanoparticles to illumination  with chiral complex fields}.   It is remarkable that in the phenomenon of dichroism on illumination with chiral light, {\it the incident reactive helicity arises in  the near-field region}, as we have shown,  while  it is well-known that  the incident optical helicity  appears  from the determination of power emitted  in the far-zone. Because of this, we call {\it reactive dichroism} the phenomenon by which this incident {\it reactive helicity becomes discriminatory} for enantiomeric separation. We propose near-field observation experiments of this reactive phenomenon.

 Given the interplay between  reactive and   radiative quantities of electromagnetic fields, we believe that the concepts studied in this work enrich  the landscape of photonics as regards nanoantennas and nanoparticle interactions with light. We expect that the observability of these reactive effects and quantities should form the basis of future experiments and techniques.  In this respect, addressing reactive quantities on higher order multipole resonance excitation will be a subject of interest for future studies.  \color{red}This is of special interest  for e.g. nanosensing advances \cite{calda} and   all-dielectric thermonanophotonics \cite{zograf}, to be used in  effective biomedical diagnosis and therapies. \color{black}

Although our analysis has been focused on the nanoscale, these results are equally valid in the microwave range due the scaling property of high index particles as Huygens sources, which remain with the same characteristics of generating large electric and magnetic dipole and multipole resonances as the illumination wavelength increases.

\section*{Acknowledgments}
MN-V work was  supported by Ministerio de Ciencia e Innovación  of Spain, grant PGC2018-095777-B-C21. 
X.X acknowledges the National Natural Science Foundation of China (11804119). Helpful comments from two anonymous referees are appreciated.

\onecolumngrid

\renewcommand{\theequation}{B-\arabic{equation}}
  \setcounter{equation}{0} 
\renewcommand{\figurename}{Figure A1}
  \setcounter{figure}{0}
\appendix

\section{PROOF OF EQS. (\ref{bcfor26a1}) and (\ref{bcfor26b1})  FOR THE ENERGY FLOW}

 Since ${\bf B}({\bf r})=(1/ik) \nabla\times {\bf E}({\bf r})$, using Eqs. (\ref{bcfor19})-(\ref{bcfor19bis}) we have
\be
{\bf B}({\bf r})=\frac{1}{k}\int_{-\infty}^{\infty}d^2{\bf K}\,\,[ {\bf k}\times{\bf e}({\bf K})] \exp [i( {\bf K}\cdot {\bf R}+ k_z z)]. \,\,\,\,\,\,\,\,\,\,\,\,  \,\,\,\,\,\,\,\,\label{bcfor19a}
\ee
Then the CPV flux across the plane $z=0$ is
\be
 \Phi^{Poynt}=
\frac{c}{8\pi}\int_{-\infty}^{\infty} d^2 {\bf R}\, [{\bf E}({\bf r})\times{\bf B}^*({\bf r})]\cdot \hat{\bm z}\,\,\,\,\,\,\,\,\,\,\,\,  \,\,\,\,\,\,\,\,\nonumber \\
 =\frac{c}{8\pi}\frac{1}{k} \int_{-\infty}^{\infty}d^2{\bf R}\,\int_{-\infty}^{\infty}d^2{\bf K}\, d^2{\bf K} ' \,
[ {\bf e}({\bf K'})\times({\bf k }^*\times{\bf e}^*({\bf K}))]\cdot \hat{\bm z}
 \exp [i({\bf K}' - {\bf K})\cdot {\bf R}+(k'_{z} - k_z^*)z]\}.\,\,\,\,\,\, \,\,\,\,\,\, \,\,\,\,\,\,\,\,\label{bcfor191}
\ee
The ${\bf R}$-integral yields  $\delta({\bf K}' - {\bf K})=(1/2\pi)^2\int_{-\infty}^{\infty}d^2{\bf R} \exp [i({\bf K}' - {\bf K})\cdot {\bf R}]$, and subsequent integration in  ${\bf K}'$ yields for the right side of (\ref {bcfor191}):
\be 
\Phi^{Poynt}=
\frac{\pi c}{2k}\int_{\infty}^{\infty}d^2{\bf K}\,[{\bf k}^*|{\bf e}({\bf K})|^2 -{\bf e}^*({\bf K})({\bf k}^*\cdot {\bf e}({\bf K}))]\cdot  \hat{\bm z}
\,\,\,\,\,\,\,\,\,\,\,\,\,\, \,\,\,\,\,\nonumber \\
=\frac{\pi c}{2k}\int_{K\leq k}d^2{\bf K}\,q_h |{\bf e}_h({\bf K})|^2
-i\frac{\pi c}{2k}\int_{K> k}d^2{\bf K}\,[q_e|{\bf e}_e({\bf K})|^2 -2q_e|{e}_{e\,z } ({\bf K})|^2].\,\,\,\,\,\,\,\,\,\,\,\,\,\,\,\,\,\,\, \label{bcfor19a1}
\ee
Where  the ${\bf K}$-integral is split into its homogeneous and evanescent parts with subscripts $h$ and $e$, respectively. Then  $ {\bf k}_h$ is the real wavevector $({\bf K}, q_h)$ of the homogeneous propagating plane wave component of complex amplitude  ${\bf e}_h({\bf K})$, whereas the complex wavevector $ {\bf k}_e=({\bf K}, iq_e)$ corresponds to  each  evanescent plane wave component of complex amplitude  ${\bf e}_h({\bf K})$. 
Also we  have taken into account that ${\bf k}_h^*\cdot {\bf e}_h({\bf K})=0$ and  ${\bf k}_e\cdot {\bf e}_e({\bf K})=0$, therefore  ${\bf k}_e^*\cdot {\bf e}_e({\bf K})=2{\bf K}\cdot{\bf e}_{e,\perp}({\bf K})=-2iq_e{ e}_{e\,z } ({\bf K})$.  [${\bf e}_e({\bf K})=({\bf e}_{e\,\perp}({\bf K}), e_{e\,z}({\bf K})$].

Taking real and imaginary parts in  (\ref{bcfor19a1}), one has
\be
\Phi^{RPoynt}= \frac{\pi c}{2k}\int_{K\leq k}d^2{\bf K}\,q_h |{\bf e}_h({\bf K})|^2,
 \label{bcfor24a1}
\ee
and
\be
\Phi^{IPoynt}=-\frac{\pi c}{2k}\int_{K> k}d^2{\bf K}\,[q_e|{\bf e}_e({\bf K})|^2 -2q_e|{ e}_{e\,z } ({\bf K})|^2]. \,\,\,\,\,\,\,\,\,\,\,\,\,\,\label{bcfor24b1}
\ee
Which are Eqs.(6) and (7) of the main text.

\renewcommand{\theequation}{B-\arabic{equation}}
  \setcounter{equation}{0} 
\renewcommand{\figurename}{Figure A1}
  \setcounter{figure}{0}

\section{PROOF OF EQS. (\ref{bcfora2011}) AND (\ref{bcfora2021a})    }

Using Eqs. (\ref{bcfor19})-(\ref{bcfor19bis}),  the integral on $z=0$ of  ${\bf E}({\bf r})\cdot{\bf B}^*({\bf r})$ is
\be
\int_{-\infty}^{\infty} d^2 {\bf R}\, [{\bf E}({\bf r})\cdot{\bf B}^*({\bf r})]
=\frac{1}{k} \int_{-\infty}^{\infty}d^2{\bf R}\,\int_{-\infty}^{\infty}d^2{\bf K}\, d^2{\bf K}' 
[ {\bf e}({\bf K'})\cdot({\bf k }^* \times{\bf e}^*({\bf K}))]
 \exp [i({\bf K}' - {\bf K})\cdot {\bf R}]\}.\,\,\,\,\,\,\,\,\,\,\,\,\,\,\, \label{bcfora19}
\ee
 And following the same procedure as in Appendix A, we are led to 
\be
\int_{-\infty}^{\infty} d^2 {\bf R}\,[{\bf E}({\bf R})\cdot{\bf B}^*({\bf R})]=
\frac{(2\pi)^2}{k}  \int_{-\infty}^{\infty}d^2{\bf K}\,\, {\bf k}^* \cdot  [{\bf e}^*({\bf K})\times{\bf e}({\bf K})] \,\,\,\,\,\,\,\,\,\,\,\,\,\,\,\,\,\,\,\,\,\,\,\,\,\,\,\,\,\,
\,\nonumber \\
=\frac{(2\pi)^2}{k}\{ \int_{K\leq k}d^2{\bf K}\, {\bf k}_h \cdot  [{\bf e}_h^*({\bf K})\times{\bf e}_h({\bf K})]
+ \int_{K> k}d^2{\bf K}\,{\bf K}\cdot  [{\bf e}^*_e({\bf K})\times{\bf e}_e({\bf K})] 
+ \int_{K> k}d^2{\bf K}\ (-i q_e)   [{\bf e}_e^*({\bf K})\times{\bf e}({\bf K})]_z\}. \,\,\,\,\,\,\,\,\,\,\,\,\,\,\,\,
 \label{bcfora20a}
\ee
We recall that  $ [{\bf e}^*({\bf K})\times{\bf e}({\bf K})]= i\Im[{\bf e}^*({\bf K})\times{\bf e}({\bf K})]=(4ki/c) \bm{\mathcal F}_{E}({\bf K})$;  where $\bm{\mathcal F}_{E}({\bf K})$ is the density of electric spin of each angular plane wave  component in ${\bf K}$-space. Then,  taking  imaginary and real parts in (\ref{bcfora20a}), and using the subindex $h$ and $e$ for homogeneous and evanescent components,  we arrive at
\be
\int_{-\infty}^{\infty} d^2 {\bf R}\,{\hel}({\bf R},0)
=(1/2k)\sqrt{\frac{\epsilon}{\mu}}\int_{-\infty}^{\infty} d^2 {\bf R}\,Im\{{\bf E}({\bf R})\cdot{\bf B}^*({\bf R})\}   
\nonumber\\
=2\frac{(2\pi)^2}{kc} \{ \int_{K\leq k}d^2{\bf K}\, {\bf k}_h \cdot  \bm{\mathcal F}_{E\,h}({\bf K}) 
+ \int_{K> k}d^2{\bf K}\,{\bf K}\cdot  \bm{\mathcal F}_{E\,e\,\perp}({\bf K})\}, \,\, \,\,\,\,\,\,\,\,\,\,\, \,\,\,\,
 \label{bcfora201}
\ee
and
\be
\int_{-\infty}^{\infty} d^2 {\bf R}\,{\qui}({\bf R},0)=(1/2k)\sqrt{\frac{\epsilon}{\mu}}\int_{-\infty}^{\infty} d^2 {\bf R}\,Re\{{\bf E}({\bf R})\cdot{\bf B}^*({\bf R})\}
=2\frac{(2\pi)^2}{kc}
 \int_{K> k}d^2{\bf K}\, q_e \, \bm{\mathcal F}_{E \,e\,z}({\bf K}). \,\,\,\,\, \,\,\,\,\,\,\,\,\,\,\,\,\,\,\,\, \label{bcfora202a}
\ee
In  Eq. (\ref{bcfora201}) $ \bm{\mathcal F}_{E\,e}=( \bm{\mathcal F}_{E\,e\,\perp}, {\cal F}_{E\,e\,z})$,  $ \bm{\mathcal F}_{E\,e\,\perp}=({\cal F}_{E \,e\,x}\,,{\cal F}_{E \,e\,y})$. Equations  (\ref{bcfora201}) and (\ref{bcfora202a}) are (\ref{bcfora2011}) and (\ref{bcfora2021a}) of the main text.

\renewcommand{\theequation}{C-\arabic{equation}}
  \setcounter{equation}{0} 
\renewcommand{\figurename}{Figure C1}
  \setcounter{figure}{0}

\section{EVANESCENT WAVE: TIME-AVERAGED QUANTITIES, ELECTRIC AND MAGNETIC IMAGINARY SPIN AND ORBITAL MOMENTA}
The real part of  ${\bf S}$  in (\ref{CPevan}) is
\be
Re\{{\bf S}\}=<{\bf S}>=\frac{c}{8\pi\mu}[\frac{K}{k}(| T_{ \perp}|^2+|T_{\parallel}|^2)\,, 2\frac{Kq}{k^2}Im\{T_{\perp}T_{\parallel}^*\}\,, 0\,] 
\exp(-2qz)
= \frac{ck}{K}[w\,,-\frac{q}{4\pi }{\hel},0].\,\,\,\,\,\,\,\,\,\,\,\,\,\,\,\,\,\,\,\,\,\,\,\,\,\,\, \label{ReCPevan}
\ee
${\bf S}$ is well-known to be associated to the optical force on a body. Considering a magnetodielectric dipolar particle placed in the air on the $ z=0$ interface,   the  real part, or energy flow density   $<{\bf S}>$, of the CPV is known to constitute a momentum of the radiation whose $x$-component, proportional to $w$,  gives rise to an $x$-force on the particle, separately  acting on  the particle electric (e) and magnetic (m) induced dipoles \cite{nieto1}.   The component $<{\bf S}>_y$ is known to produce a lateral force proportional to $-\hel$ along $OY$ \cite{bliokh1},   due to  the interference of its e and m dipoles \cite{nieto1}. 

On the other hand, the time-averages of  the density of   spin angular momentum, $ <\bm{\mathcal S}>=(1/4\pi c^2)\bm{\mathcal F} =(1/2)(<\bm{\mathcal S}_e>+<\bm{\mathcal S}_m>)$, where $<\bm{\mathcal S}_e>=(1/8\pi kc)\Im\{{\bf E}^*\times{\bf E}\}$ and 
 $<\bm{\mathcal S}_m>=(1/8\pi kc)Im\{{\bf B}^*\times{\bf B}\}$, and of spin curl, or Belinfante spin momentum,  $<{\bf P}^S>=(1/2)[<{\bf P}_e^S>+<{\bf P}_m^S>]=(1/2)\nabla\times <\bm{\mathcal S}>$, are
\be
<\bm{\mathcal S}>= \frac{1}{8\pi c}[\frac{-2K}{k^2}Im\{T_{\perp}T^{*}_{\parallel}\}, 
-\frac{Kq}{k^3}(| T_{ \perp}|^2+|T_{\parallel}|^2)\,, 0]  
\exp(-2qz)=\frac{1}{4\pi cK}\,[k\hel\,,-\frac{4\pi q}{k }w,\,0] . \,\,\,\,\,\,\,\,\,\,\,\,\,\,\,\,\,\label{spinevan}
\ee
With $<{\bf P}_e^S>=(1/2)\nabla\times <\bm{\mathcal S}_e>$, $<{\bf P}_m^S>=(1/2)\nabla\times <\bm{\mathcal S}_m>$, and $<{\bf P}^S>$ being:
\be
<{\bf P}_e^{S}>=\frac{1}{8\pi c}[-\frac{2K q^2}{k^3}| T_{ \parallel}|^2\, 
\frac{2Kq}{k^2}Im\{T_{\perp}T^{*}_{\parallel}\} ,\,0]
\exp(-2qz) .\,\,\,\,\,\,\, \,\,\,\, \,\,\,\,\,\,\,\,\,\, \,\,\,\, \,\,\,\label{rotspinevan_e} 
\\
<{\bf P}_m^{S}>=\frac{1}{8\pi c}[-\frac{2K q^2}{k^3}| T_{ \perp}|^2\,, 
\frac{2Kq}{k^2}Im\{T_{\perp}T^{*}_{\parallel}\} ,\,0]
\exp(-2qz) .\,\,\,\,\,\,\, \,\,\,\, \,\,\,\,\,\,\,\,\,\, \,\,\,\, \,\,\,\label{rotspinevan_m} 
\\
<{\bf P}^{S}>=\frac{1}{8\pi c}[-\frac{K q^2}{k^3}(| T_{ \perp}|^2+|T_{\parallel}|^2)
\frac{2Kq}{k^2}\Im\{T_{\perp}T^{*}_{\parallel}\} ,\,0]\exp(-2qz) 
=-\frac{q}{cK}\,[\frac{q}{k }w, \,\frac{k}{4\pi}\hel,\,0] . \,\,\,\,\,\,\,\,\,\,\,\,\,\,\,\,\, \label{rotspinevan}
\ee
Equation (\ref{spinevan}) remarks that, in agreement with the second term of the right side of (\ref{bcfora2011}),  $\hel$ appears in the projection of the density of spin momentum $<\bm{\mathcal S}>$ on the $x$-direction, which is that of propagation of the evanescent wave \cite{comment}. On the other hand, (\ref{rotspinevan}) highlights the transversal $y$-component of the spin momentum proportional to $\hel$, as well as its longitudinal component, along $OX$, proportional to $w$.

Since the time-average electromagnetic field momentum density $<{\bf g}>=<{\bf S}>/c^2$ holds
\be
<{\bf g}>=<{\bf P}^O> +< {\bf P}^S>; \label{orbplusspin}
\ee
$<{\bf P}^O>$ being the density of time-averaged orbital momentum, 
one sees  from (\ref{ReCPevan}) and (\ref{rotspinevan}) that {\it the transverse $y$-component of $<{\bf g}>$  comes from  $<{P}_y^S>$, which is characterized by $\hel$}.  Both  $<{ g}_y>$  and   $<{ P}^S_y>$ are   $(-1/4\pi c)(kq/K){\hel}\exp(-2qz)$. Therefore {\it the trensverse $y$-component $<{g}_y>$ of the field momentum is provided by the transverse $y$-component  $<{ P}_y^S>$ of Belinfante's momentum}, in agreement with \cite{bliokh1}.   

In addition, from (\ref{orbplusspin}), (\ref{rotspinevan}) and   (\ref{ReCPevan}) we have for the $i$th Cartesian component of the time-averaged  orbital
momentum  density,
\be
<{P}_i^O>=(1/2)(<{ P}_{e\,i}^O>+<{ P}_{m\,i}^O>)=
(1/2)(1/8\pi kc)[Im\{E_{j}^* \partial_i E_j\}+Im\{B_{j}^* \partial_i B_j\}],\,\,\, (i,j=x,y,z), \nonumber
\ee
\be
<{\bf P}^{O}>
=\frac{1}{8\pi c}[\frac{K^3}{k^3}(| T_{ \perp}|^2+|T_{\parallel}|^2)\,, \,0\,,\,0]\exp(-2qz)
=\frac{w}{c}\,[\frac{K}{k} \, ,0\, ,0 ] . \,\,\,\,\,\,\,\,\,\,\,\,\,\,\,\,\,\,\,\,\, \,\,\,\,\label{orbitevan}
\ee
With the electric and magnetic orbital momenta:
\be
<{\bf P}_{e}^{O}>
=\frac{1}{8\pi c}[\frac{K(2K^2-k^2)}{k^3}| T_{ \parallel}|^2+\frac{K}{k}|T_{\perp}|^2\,, \,0\,,\,0]
\times\exp(-2qz), \,\,\,\,\,\,\,\,\,\,\,\,\,\,\,\,\,\,\,\,\, \,\,\,\,\label{orbitevan1}
\ee
and
\be
<{\bf P}_{m}^{O}>
=\frac{1}{8\pi c}[\frac{K(2K^2-k^2)}{k^3}| T_{\perp }|^2+\frac{K}{k}|T_{\parallel}|^2\,, \,0\,,\,0]
\times\exp(-2qz), \,\,\,\,\,\,\,\,\,\,\,\,\,\,\,\,\,\,\,\,\, \,\,\,\,\label{orbitevan2}
\ee
respectively.  $<{\bf P}^{O}>$ contains the energy density $w$ of the wave and, as such, points in the propagation 
direction $OX$ of the wave, like both $<{g}_x>$   and  the $x$-component, $K>k$, of its propagation wavevector. 
This, in agreement with \cite{bliokh1}, confers to the evanescent wave a superluminal  group velocity; although of 
course the time-average energy flow $<{ S}_x>$ propagates with speed less than $c$. Hence $<{ P}_x^{O}>$  
pushes the aforementioned small particle,  placed in the air on the interface, as radiation pressure  along the    $x$-
propagation direction.

\color{red}   On the other hand, the electric and magnetic imaginary momenta of (\ref{Imoms}) are:
\be
({\bf P}_e^{S\,I})_i=\frac{1}{8\pi kc }Re\{\partial_j ({ E}_i^{*}  { E}_j)\}\,,\,\,\,\nonumber \\
({\bf P}_m^{S\,I})_i=\frac{1}{8\pi kc}Re\{\partial_j ({ B}_i ^{*} { B}_j)\}\,, \nonumber\\
({\bf P}_e^{O\,I})_i=-\frac{1}{8\pi kc }\partial_j \frac{1}{2}\delta_{ij}|{\bf E} |^2=-\frac{1}{8\pi kc } \frac{1}{2}\partial_i|{\bf E} |^2    \,\,\,\,\,\,\,\,\,\,\,\,\,\,\,\, \,  \nonumber \\
({\bf P}_m^{O\,I})_i=-\frac{1}{8\pi kc } \frac{1}{2}\partial_i|{\bf B} |^2 \,.\,\,\,\,\,\,(i,j=x,y,z).\,\,\,\,\,\,\,\,\,\,\,\,\,\,\,  \label{bcfor5A5}
\ee
\be
{\bf P}_e^{S\,I}=\frac{1}{4\pi k^2 c}(0, -Kq\, Re\{T_{\perp}T_{\parallel}^*\},-\frac{K^2 q}{k}|T_{\parallel}|^2)\exp(-2qz).  \,\,\,\,\,\,\,\,\,\,   \label{bcfor5A5}  \label{evanreener1}
\ee
\be
{\bf P}_m^{S\,I}=\frac{1}{4\pi k^2 c}(0, \,Kq\, Re\{T_{\perp}T_{\parallel}^*\},-\frac{K^2 q}{k}|T_{\perp}|^2)\exp(-2qz).  \,\,\,\,\,\,\,\,\,\,  \label{bcfor5A5}  \label{evanreener2}
\ee
\be
{\bf P}_e^{O\,I}=\frac{q}{8\pi k c}(0, 0,  \frac{K^2 +q^2}{k^2}|T_{\parallel}|^2+|T_{\perp}|^2)\exp(-2qz).    \,\,\,\,\,\,\,\,\,\,\,\,\,\,\,\,   \label{bcfor5A5}\label{evanreener3}
\ee
\be
{\bf P}_m^{O\,I}=\frac{q}{8\pi k c}(0, 0,  \frac{K^2 +q^2}{k^2}|T_{\perp}|^2+|T_{\parallel}|^2)\exp(-2qz).    \,\,\,\,\,\,\,\,\,\,\,\,\,\,\,\,   \label{bcfor5A5}\label{evanreener4}
\ee
\color{black}

\renewcommand{\theequation}{D-\arabic{equation}}
  \setcounter{equation}{0} 
\renewcommand{\figurename}{Figure D1}
  \setcounter{figure}{0}

\section{DIPOLE FIELDS}
We express the fields  emitted by a dipole with electric and magnetic moments ${\bf p}$ and ${\bf m}$ as \cite{jackson}, 
\be
{\bf E}^{(s)}({\bf r})=\{\frac{k^2}{\epsilon r}[\hat{\bm r}\times({\bf p}\times \hat{\bm r} )]+\frac{1
}{\epsilon}
[3\hat{\bm r} (\hat{\bm r} \cdot{\bf p})-{\bf p}](\frac{1}{r^3} 
-\frac{ik}{r^2})   
-\sqrt{\frac{\mu}{\epsilon}}(\hat{\bm r} \times{\bf m})(\frac{k^2}{r} +\frac{ik}{r^2})\}e^{ikr}. \,\,\,\, ({\bf r}=r\hat{\bm r}). \label{app1}\,\,\,\,\, \,\,\,\,\,\,\,\,\,\,\,\,\, \,\,\,\,\,\,\,\, \,\,\,\,\, \,\,\,\,\,\,\,\,  \\ 
{\bf B}^{(s)}({\bf r})=\{\frac{\mu k^2}{r}[\hat{\bm r} \times({\bf m}\times \hat{\bm r} )]+{\mu}[3\hat{\bm r} (\hat{\bm r} \cdot{\bf m})-{\bf m}](\frac{1}{r^3}-\frac{ik}{r^2})   
+\sqrt{\frac{\mu}{\epsilon}}(\hat{\bm r} \times{\bf p})(\frac{k^2}{r} +\frac{ik}{r^2})e^{ikr}. \,\,\,\,\, \,\,\,\,\,\,\, \,\,\,\,\,\,\,\, \,\,\,\,\, \,\,\,\,\,\,\,\, \,\,\, \,\,\,\,\, \,\,\,\,\,\,\,\,\,\,\, \,\,\,\,\, \,\,\,\,\,\,\,\,\label{app2}
\ee

\renewcommand{\theequation}{E-\arabic{equation}}
  \setcounter{equation}{0} 
\renewcommand{\figurename}{Figure E1}
  \setcounter{figure}{0}

\section{PROOF OF  EQS. (\ref{Wwtot1}) AND (\ref{Wtot1})}
The reactive power $W_{react}^{(s)}=\frac{ck}{3r^3}(|{\bf p}|^2-|{\bf m}|^2)$ may also be obtained  by performing the volume integration  of the right side of (\ref{Wwtot1}) for the magnetoelectric dipole:
 \be
<\tilde{W}_e^{(s)}>=\int_{V_{\infty}-V}d^3r <\tilde{w}_e^{(s)}>
=\frac{1}{16\pi}\int_r^{\infty}drr^2\int_{0}^{2\pi}d\phi\int_{0}^{\pi} d\theta\sin\theta
 [|{\bf E}^{(s)}|^2-|{\bf E}_{FF}^{(s)}|^2]
=\frac{k^3}{6}[|{\bf p}|^2 (\frac{1}{(kr)^3}+\frac{1}{kr}) +\frac{|{\bf m}|^2}{kr}] , \,\,\,\,\,\,\,\,\,\,\,\,\,\,\,\,\label{ste} 
\ee
 and
\be
<\tilde{W}_m^{(s)}>=\int_{V_{\infty}-V}d^3r <\tilde{w}_m^{(s)}>
=\frac{1}{16\pi}\int_r^{\infty}drr^2\int_{0}^{2\pi}d\phi\int_{0}^{\pi} d\theta\sin\theta  
 [|{\bf B}^{(s)}|^2-|{\bf B}_{FF}^{(s)}|^2] 
=\frac{k^3}{6}[|{\bf m}|^2 (\frac{1}{(kr)^3}+\frac{1}{kr}) +\frac{|{\bf p}|^2}{kr}]. \,\,\,\,\,\,\,\,\,\,\,\,\,\,\,\, \label{stm} 
\ee
Having used Eqs. (\ref{app1}) and (\ref{app2}) of Appendix D, and ${\bf E}_{FF}^{(s)}$  and ${\bf B}_{FF}^{(s)}$ being the radiated  far-fields with $r^{-1}$ dependence. Thus the time-averaged reactive power is
\be
2\omega(<\tilde{W}_m^{(s)}>-<\tilde{W}_e^{(s)}>)=W_{react}^{(s)}=\frac{ck}{3r^3}(|{\bf m}|^2-|{\bf p}|^2). \,\,\,\,\,\,\,\,\,\,\,\,\,\,\,\, \label{stme}
\ee
Which is (\ref{Wwtot1}).  

The  total time-averaged  energy outside the volume $V$ is after Eqs. (\ref{ste}) and (\ref{stm}),
\be
<\tilde{W}_T^{(s)}>=<\tilde{W}_e^{(s)}>+<\tilde{W}_m^{(s)}>
=\frac{1}{6}(|{\bf p}|^2+|{\bf m}|^2) (\frac{1}{r^3}+\frac{2k^2}{r}). \label{Wtot2}
\ee
Which is (\ref{Wtot1}) when $r\rightarrow a$ .

\renewcommand{\theequation}{E-\arabic{equation}}
  \setcounter{equation}{0} 
\renewcommand{\figurename}{Figure F1}
  \setcounter{figure}{0}

\twocolumngrid


\begin{thebibliography}{99}

\bibitem{raether} H. Raether, {\it Surface Plasmons on Smooth and Rough Surfaces and on Gratings}, Springer-Verlag (Berlin, 1988).

\bibitem{kolokolov} A. A. Kolokolov and G. V. Skrotskii, Interference of reactive components of an electromagnetic field, Sov. Phys. Usp. {\bf 35}, 1089 1093 (1992).
\bibitem{harrington} R. F. Harrington, {\it Time-harmonic Electromganetic Fields}, J. Wiley (New York, 2001).
\bibitem{stratton} J. A. Stratton,{\it  Electromagnetic Theory}, Mc Graw-Hill, (New York, 1941).
\bibitem{wheeler} H. A. Wheeler, Fundamental limitations of small antennas, Proc. IRE,  {\bf 35}, 1479-1484  (1947).
\bibitem{chu} L. J. Chu, Physical limitations on omni-directional antennas, J. Appl. Phys., {\bf 19}, 1163-1175 (1948).
\bibitem{collin} R. E. Collin and S . Rothschild, Evaluation of antenna $Q$,  IEEE Trans. Antennas Propagat., {\bf AP-12}, 23-21 (1964); R. E. Collin, Minimum $Q$ of small Antennas,  J. Electromag. Waves and Appl., {\bf 12} 1369-1393 (1998) DOI: 10.1163/156939398X01457.

\bibitem{jackson} J.D. Jackson, {\it Classical Electrodynamics},  2nd edn. J. Wiley (New York, 1975).
\bibitem{mcLean} J. S. McLean,  A Re-Examination of the Fundamental Limits on the Radiation Q of Electrically Small Antennas,     IEEE Trans. Antennas Propag. {\bf AP-44},  672-676  (1996). 

\bibitem{geyi} W. Geyi and  P. Jarmuszewski,  The Foster Reactance Theorem for Antennas and Radiation Q, IEEE Trans. Antenn. Propag.   {\bf 48}, 401-408 (2000);  W. Geyi, Foundatios of applied electrodynamics, J. Wiley, (New York, 2010). Sec. 4.4.1.

\bibitem{balanis} C.A. Balanis, {\it Antenna Theory}, 4th ediion, J. Wiley, (New York, 2016).
\bibitem{alu} C. A. Valagiannopoulos and A. Alu, "The Role of Reactive Energy in the Radiation
by a Dipole Antenna".  IEEE Trans. Antennas  Propag.  {\bf 63}, 3736-3741 (2015).
\bibitem{tang}  Y. Tang and A. E.  Cohen. Optical chirality and its interaction with matter. Phys. Rev. Lett. {\bf 104}, 163901 (2010).
\bibitem{barnett2} R. P. Cameron, S. M. Barnett  and A. M. Yao,   Optical helicity, optical spin and related quantities in electromagnetic theory, New. J. Phys. {\bf 14}, 053050 (2012). 
\bibitem{corbato} I. Fernandez-Corbaton, I. and G. Molina-Terriza, Role of duality symmetry in transformation optics. Phys. Rev. B {\bf 88}, 085111 (2013).
\bibitem{nietoheli}M.  Nieto-Vesperinas,   Optical theorem for the conservation of electromagnetic helicity: significance for molecular energy transfer and enantiomeric discrimination by circular dichroism, Phys. Rev. A {\bf 92}, 023813 (2015; M. Nieto-Vesperinas, Chiral optical fields: a unified formulation of helicity scattered from particles and dichroism enhancement, Phil. Trans. R. Soc. A {\bf 375}, 20160314 (2017).
\bibitem{banzer} S. Nechayev and P. Banzer, Mimicking chiral light-matter interaction, Phys. Rev. B {\bf 99}, 241101(R) (2019).

\bibitem{barnett1} F. Crimin, N. Mackinnon, J. B. Götte and S. M. Barnett, Optical helicity and chirality: Conservation and sources, Appl. Sci.  {\bf 9}, 828  (2019). doi:10.3390/app9050828.

\bibitem{yan} S. Yan, M. Li, Y. Liang, Y. Cai and B. Yao, Spin momentum-dependent orbital motion, New J. Phys. {\bf 22}, 053009  (2020).

\bibitem{nieto1} M. Nieto-Vesperinas,  J. J. Saenz, R. Gomez-Medina and L. Chantada, Optical forces on small magnetodielectric particles. Opt. Express {\bf  18}, 11428–11443 (2010).
\bibitem{bliokh1} K. Y. Bliokh, A. Y. Bekshaev and F. Nori, Extraordinary momentum and spin in evanescent waves, Nat. Comm.
{\bf 5} 3300 (2014);  M. Antognozzi, C. R. Bermingham, R. L. Harniman, S. Simpson, J. Senior, R. Hayward, H. Hoerber, M. R. Dennis, A. Y. Bekshaev, K. Y. Bliokh and  F. Nori, Direct measurements of the extraordinary optical momentum and transverse spin-dependent force using a nano-cantilever, Nat. Phys. {\bf 12},  731–735 (2016).
\bibitem{bliokh_rep} K. Y. Bliokh and F. Nori, Transverse and Longitudinal Angular Momenta of Light, Phys. Rep. {\bf 592}, 1 (2015).
\bibitem{bliokh2} K. Y. Bliokh, Y. S. Kivshar, and F. Nori, Magnetoelectric Effects in Local Light-Matter Interactions, Phys. Rev. Lett. {\bf 113}, 033601 (2014).

\bibitem{xu} X. Xu and  M. Nieto Vesperinas,  Azimuthal imaginary Poynting momentum density, Phys. Rev. Lett.
{\bf123}, 233902 (2019).

\bibitem{xiao} X. Xu, M. Nieto-Vesperinas, C.-W. Qiu, X. Liu, D. Gao, Y. Zhang, and B. Li, Kerker-Type Intensity-Gradient Force of Light, Laser Photonics Rev. {\bf 14}, 1900265 (2020).


\bibitem{novotny1}  P. Bharadwaj, B. Deutsch, and L. Novotny, Optical Antennas, Advan.  Opt. and Photon. {\bf 1}, 438–483 (2009); L. Novotny and N. van Hulst, Antennas for light, Nat. Photon.  {\bf 5}, 83 (2011).

\bibitem{norris1}  L. V. Poulikakos, P. Thureja, A. Stollmann, E. De Leo and D. J. Norris, Chiral Light Design and Detection Inspired by Optical Antenna Theory.  Nano Lett. {\bf 18}, 4633-4640 (2018).

\bibitem{barnes} W. L. Barnes and  S. A. R. Horsley, Classical antennae, quantum emitters, and densities of optical states, arXiv:1909.05619 (2019).
\bibitem{hecht} P. Biagioni, J.-S. Huang and B. Hecht, Nanoantennas for visible and infrared radiation, Rep. Prog. Phys. {\bf 75}, 024402 (2012) .

\bibitem{ziolkowski} I. Liberal, I. Ederra, R. Gonzalo and R. W. Ziolkowski, Induction Theorem Analysis of Resonant Nanoparticles: Design of a Huygens Source Nanoparticle Laser, Phys. Rev. Appl. {\bf 1},  044002 (2014).

\bibitem{engheta} I. Liberal and N. Engheta, Nonradiating and radiating modes excited by quantum emitters in open epsilon-near-zero cavities, Sci. Adv. {\bf 2}, e1600987 (2016).

\bibitem{saad}  M. S. Bin-Alam, O. Reshef,  Y. Mamchur, M. Z. Alam, G. Carlow, J. Upham, B. T. Sullivan, 
J-M. Menard, M. J. Huttunen, R. W. Boyd, and K. Dolgaleva,  Ultra-high-Q resonances in plasmonic metasurfaces,
arXiv:2004.05202 (2020).

\bibitem{won} R. Won, Into the Mie-tronic era, Nat. Photon. {\bf 13}, 585–587 (2019).
\bibitem{bonod} N. Bonod and Y. Kivshar, All-dielectricMie-resonant metaphotonics, Compt. Rendus Phys. https://doi.org/10.5802/crphys.31    (2020).

\bibitem{nietoSi} A. Garcia-Etxarri, R. Gomez-Medina, L. S. Froufe-Perez, C. Lopez, L. Chantada,  F. Scheffold,  J. Aizpurua, M. Nieto-Vesperinas and J. J. Saenz, Strong magnetic response of submicron Silicon particles in the infrared, arXiv:1005.5446v1, 29 May 2010; Opt. Express {\bf 19} 4816 (2011).

\bibitem{nieto2011}M. Nieto-Vesperinas, R. Gomez-Medina and J. J. Saenz, Angle suppressed
scattering and optical forces on submicrometer dielectric particles. , J. Opt. Soc. Am. A {\bf 28}, 54 (2011).

\bibitem{mlight} A. I. Kuznetsov, A. E. Miroshnichenko, Y. H. Fu, J. B. Zhang and  B. Luk’yanchuk, Magnetic light, Sci. Reps. {\bf 2},  492 (2012).

\bibitem{staude} M. Decker and I. Staude, Resonant dielectric nanostructures: a low-loss
platform for functional nanophotonics, J. Opt. {\bf 18},  103001 (2016); I. Staude, T. Pertsch and Y. S. Kivshar, 
All-dielectric resonant Meta-Optics lightens up, ACS Photonics {\bf 6}, 802-814 (2019).

\bibitem{kivshar} A. I. Kuznetsov, A. E. Miroshnichenko, M. L. Brongersma, Y. S. Kivshar, B. Luk'yanchuk, Optically resonant dielectric nanostructures, Science {\bf  354}, 2472 (2016).
\bibitem{nietolibrev} M. Nieto-Vesperinas, Fundamentals of Mie scattering. In {\it Dielectric Metamaterials:
Fundamentals, Designs, and Applications}, I. Brener, S. Liu, I. Staude, J. Valentine and C. Holloway, eds.  Chapt. 2. Elsevier (Amsterdam, 2019). R. Paniagua-Dominguez, B. Luk’yanchuk and A. I. Kuznetsov, Control of scattering by isolated dielectric nanoantennas. {\it Ibid.}, Chapt. 3.

\bibitem{comment0}  The helicity of  time-harmonic wavefields  is proportional  to their chirality \cite{tang}. Both quantities differ only by the square of the wavenumber \cite{nietoheli}.

\bibitem{kamenetskii}  E. O. Kamenetskii,  M. Berezin and  R. Shavit, Microwave magnetoelectric fields: helicities and reactive power flows, Appl. Phys. B {\bf 121},  31–47 (2015).

\bibitem{kerker} M. Kerker, D. S. Wang, and C. L. Giles, Electromagnetic scattering by magnetic spheres, J. Opt. Soc. Am. {\bf 73 },
765 (1983).


\bibitem{nietoJNano} R. Gomez-Medina, B. Garcia-Camara, I. Suarez-Lacalle, F. Gonzalez,
F. Moreno, M. Nieto-Vesperinas and J. J. Saenz,  Electric and magnetic dipolar response of germanium
nanospheres: interference effects, scattering anisotropy, and optical forces, J. Nanophotonics {\bf 5}, 053512  (2011).

\bibitem{geffrin} J. M. Geffrin, B. Garcia-Camara, R. Gomez-Medina, P. Albella, L., S. Froufe-Perez, C. Eyraud, A. Litman, R. Vaillon, F. Gonzalez, M. Nieto-Vesperinas, J. J. Saenz and F. Moreno, Magnetic and electric coherence in
forward- and back-scattered electromagnetic waves by a single dielectric subwavelength sphere,
Nat. Comm. {\bf3}, 1171 (2012).

\bibitem{lapin} S. Person, M. Jain, Z. Lapin, J. J. Saenz, G. Wicks and L. Novotny, Demonstration of zero optical backscattering from single nanoparticles,  Nano Lett. {\bf 13}, 1806 (2013). 

\bibitem{banzer1} A. Bag, M. Neugebauer, P. Woźniak, G. Leuchs, and P. Banzer, Transverse Kerker Scattering for Angstrom Localization of Nanoparticles, Phys. Rev. Lett. {\bf 121}, 193902 (2018).

\bibitem{olmos} J. Olmos-Trigo, C. Sanz-Fernandez, D. R. Abujetas, J. Lasa-Alonso, N. de Sousa, A. García-Etxarri, J. A. Sanchez-Gil, G. Molina-Terriza and J. J. Saenz, Kerker Conditions upon Lossless, Absorption, and Optical Gain Regimes, Phys. Rev. Lett. {\bf 125}, 073205 (2020); J. Olmos-Trigo, D. R. Abujetas, C. Sanz-Fernández, J. A. Sánchez-Gil and J. J. Saenz, Optimal backward light scattering by dipolar particles, 
Phys. Rev. Research {\bf 2}, 013225  (2020);
 J. Olmos-Trigo, D. R. Abujetas, C. Sanz-Fernandez,  X. Zambrana-Puyalto, N. de Sousa, J. A. Sanchez-Gil, and J. J. Saenz, Unveiling dipolar spectral regimes of large dielectric Mie spheres from helicity conservation, Phys. Rev. Research {\bf 2}, 043021  (2020).
\color{red}
\bibitem{note1} If the embedding medium is lossy and hence it also  presents a certain dispersion, or complex, there will be  terms associated with effects additional to those adressed in this research. See e.g. Section 1.8 of \cite{harrington} for the complex Poynting theorem and \cite{poulikakos1} for the conservation of optical chiraliity. For instance, for lossy dispersive media the electric and magnetic energy terms should be of the form $(1/16\pi)\partial_{\omega}[\epsilon'(\omega)+i\epsilon''(\omega)]|{\bf E}|^2$ and $(1/16\pi)\partial_{\omega}[\mu'(\omega)+i\mu''(\omega)]|{\bf H}|^2$, [cf. e.g. Section 80 of  L. D. Landau and E.M. Lifshitz, {\it Electrodynamics of Continuous Media}, Pergamon Press, (Oxford,  1984)], which is in fact an approximation for time slowly-varying fields. Such generalizations, and their related problems \cite{geyi,ziolkowski} are outside the aims of this work.

\color{black}
\bibitem{mandel} L. Mandel and E. Wolf,{\it  Optical Coherence and Quantum Optics}, Cambridge U.P., (Cambridge, 1995).
\bibitem{nietolibro} M. Nieto-Vesperinas, {\it Scattering and Diffraction in Physical Optics}, 2nd edition, World Scientific, (Singapore, 2006).

\bibitem{bliokh4}  K. Y. Bliokh, A. Y. Bekshaev and F. Nori, Dual electromagnetism: helicity, spin, momentum and angular momentum, New J. Phys. {\bf 15},  033026 (2013).
\bibitem{poulikakos1} L. V. Poulikakos,  P. Gutsche, K. M. McPeak,  S. Burger, J. Niegemann, C. Hafner and D. J. Norris, Optical chirality flux as a useful far-field probe of chiral near fields, ACS Photonics {\bf 3}, 1619–1625 (2016).\bibitem{poulikakos2} P. Gutsche, L. V.  Poulikakos,  M. Hammerschmidt, S.  Burger and F. Schmidt, Time-harmonic optical chirality in inhomogeneous, space. Proc. SPIE 9756, 97560X arXiv:1603.05011 (2016).

\bibitem{gutsche} P.  Gutsche and  M. Nieto-Vesperinas, Optical Chirality of Time-Harmonic Wavefields for Classification of Scatterers. Sci. Reps. {\bf 8}, 9416 (2018).

\bibitem{norris_use} Notice the difference of  \color{red} Eqs. (\ref{bcpoy53})-(\ref{bcpoy63}) with the    conservation equation and quantities of  \cite{poulikakos1}\color{red}, (cf. Eqs. (S.I.12)-(S.I.15) of the Supporting Information of \cite{poulikakos1}).  In  homogeneous lossless media addressed here, (S.I.12)-(S.I.15) describe purely real quantities and  $\chi_e -\chi_m=0$, so that the conservation equation (S.I.12) of \cite{poulikakos1} does not provide a law for reactive quantities, although  both the optical chirality flux  (S.I.15)  and the real conservation equation (S.I.12)  are $k^2$ times the flow of helicity (\ref{helflow}) and  the conservation of  helicity (\ref{bcpoyReal}), respectively.  

In order to get complex expresions  (S.I.12)-(S.I.15), and  $Re[\chi_e -\chi_m]\neq 0$,  $Im[\chi_e -\chi_m]\neq 0$),   lossy media should be considered, but then  the imaginary parts of (S.I.12)-(S.I.15) are different to (\ref{bcpoy53})-(\ref{bcpoy63}).  Hence,  the  imaginary part  (\ref{bcpoy53}) of the complex conservation equation  (\ref{bcpoy63}) is  a novel law for the reactive helicity ${\qui}$ and its flow ${\cal F}_\qui$ .

\bibitem{note2} 
Let  ${\bf E}=(e^{+}{\bm \epsilon}^{+}+ e^{-}{\bm \epsilon}^{-})\cos kz$,  ${\bf  B}=(e^{+}{\bm \epsilon}^{+}- e^{-}{\bm \epsilon}^{-})\sin kz$,  be an  elliptically polarized standing wavefield, expressed in the helicity basis $\hat{\bm\epsilon}^{\pm}=\frac{1}{\sqrt{2}}(\hat{\bm x}\pm i\hat{\bm y})$, with amplitudes CPL(+) (left circularly polarized)   $e^{+}$ and CPL(-) (right circularly polarized)  $e^{-}$. One has ${\bf E}\times {\bf B}^*=-(i/2)(|(e^{+}|^2+|e^{-}|^2)\sin 2kz\, \hat{\bm z}$, \,  $|{\bf B}|^2-|{\bf E}|^2=-(|(e^{+}|^2+ |e^{-}|^2)\cos 2kz$, and hence $\nabla\cdot\Im\{{\bf S}\}=2\omega (<w_m> -<w_e>)$. In addition,
 ${\bf E}\cdot {\bf B}^*=(1/2)(|(e^{+}|^2- |e^{-}|^2)\sin2kz$,  ${\bf B}^*\times {\bf B}-{\bf E}^*\times {\bf E}=-i(|(e^{+}|^2-|e^{-}|^2)\cos 2kz\, \hat{\bm z}$, so that
$\nabla\cdot{\cal F}_\qui=2\omega \qui$.  

Therefore  this kind of wave has no time-averaged energy transport $<{\bf S}>$, nor helicity density ${\hel}$, but their densities of reactive power $2\omega(<w_m>-<w_e>)$,  IPV,  reactive helicity  $\qui$ and its flow ${\cal F}_\qui$ , are non-zero. Note that if the standing wave is linearly polarized, (e.g. $E_y=B_x=0$), it has IPV and stores reactive power, 
but $\qui=0$. \color{black}

\color{red}
\bibitem{nieto_torque} M. Nieto-Vesperinas, Optical torque: Electromagnetic spin and orbital-angular-momentum conservation laws and their significance.Phys. Rev. A {\bf 92}, 043843 (2015).
\color{black}

\bibitem{born} M. Born and E. Wolf,{\it  Principles of Optics}, Cambridge University Press, Cambridge (1995).
\bibitem{nietoOL} M. Nieto-Vesperinas and J. J. Saenz, “Optical forces from an
evanescent wave on a magnetodielectric small particle,” Opt. Lett. {\bf 35}, 4078–4080 (2010).

\bibitem{neugebauer}M. Neugebauer, T. Bauer, A. Aiello, and P. Banzer, Measuring the Transverse Spin Density of Light, Phys.Rev. Lett. {\bf 114}, 063901 (2015).

\bibitem{comment} Notice that Eqs.(\ref{evan}) correspond to a wavefield ${\bf E}({\bf r})$ with  angular spectrum ${\bf e}({\bf K'})= \left(-\frac{k'_z}{k } T_{ \parallel}, T_{\perp},
\frac{K'}{k } T_{\parallel}\right) \delta({\bf K}' - {\bf K}), ({\bf k}=({\bf K},k_z), {\bf K} =(K_x,0), k_z=iq, K_x>k)$.
\bibitem{chaumetPRB} P. C. Chaumet and M. Nieto-Vesperinas,  Coupled dipole method determination of
the electromagnetic force on a particle over a flat dielectric substrate. Phys. Rev. B {\bf  61}, 14119-14127 (2000).

\bibitem{noteinterf}  Note that  $<\tilde{w}_e^{(s)}>$ and $<\tilde{w}_m^{(s)}>$ are 
not the electric and magnetic energy densities of only the near plus intermediate fields, say ${\bf E}^{(NF,IF)}$, ${\bf B}^{(NF,IF)}$, since they also contain an interference term  of these fields  with the  far fields:
 $2 Re \{{\bf E}^{(NF,IF)}\cdot{\bf E}^{(FF)}\}$ and $ 2 Re\{{\bf B}^{(NF,-F)}\cdot{\bf B}^{(FF)}\}$,  respectively.

\bibitem{bohren} C. F. Bohren and D. R. Huffman, {\it Absorption and Scattering of Light by Small Particles}, (John Wiley and Sons, 1998).

\bibitem{kivsharmeta} K. Koshelev and Y. Kivshar, Dielectric Resonant Metaphotonics, ACS Photonics. https://pubs.acs.org/doi/pdf/10.1021/acsphotonics.0c01315.
\bibitem{kivsharBIC} M. V. Rybin, K. L. Koshelev, Z. F. Sadrieva, K. B. Samusev, A. A. Bogdanov,
M. F. Limonov and Y. S. Kivshar,  High-Q Supercavity Modes in Subwavelength Dielectric Resonators. Phys Rev. Lett. {\bf 119}, 243901 (2017).

\color{red}
\bibitem{bornseries} Note that one could address a medium composed of particles, in  which transport of light occurs.  The existence of multiple scattering may be associated with a slow convergence of the Born series due to some coupling between the particles, stringent conditions for convergence under excitation of their resonances, (see e.g. N. A. Ustimenko, D. F. Kornovan, K. V. Baryshnikova, A. B. Evlyukhin, and M.  I. Petrov, Multipole Born series approach to light scattering by Mie-resonant nanoparticle structures,  arXiv:2108.11920v1 26 Aug 2021), or even with no convergence at all if there is either strong coupling between the induced dipoles (or multipoles). The latter two effects should be associated with higher Q-factors,  and thus with larger amounts of stored and reactive powers. The same we would expect to rule the behavior of the scattered helicity and a quality factor which might be introduced for this quantity; this being an area of possible future exploration.
\color{black}


\bibitem{comment_CPL} For a circularly polarized incident plane wave of unit amplitude, the induced dipoles are 
${\bf p}=\alpha_e{\bf E}^{(i)}$ and ${\bf m}=\alpha_m{\bf B}^{(i)}$ with  ${\bf E}^{(i)}=\hat{\bm  \epsilon}
^{\pm}$, ${\bf B}^{(i)}=\mp  i\hat{\bm  \epsilon}^{\pm}$,  $\hat{\bm  \epsilon}^{\pm}=\frac{1}{\sqrt{2}}(\hat{\bm x}\pm i\hat{\bm y})$,  the upper and lower sign  applies according to whether it is left circular, CPL(+), or right circular, CPL(-), respectively. 
Since at K1:  $\alpha_m=\alpha_e$, then  ${\bf p}=\alpha_e \hat{\bm \epsilon}^{\pm}$ and ${\bf m}=\mp i\alpha_e 
\hat{\bm  \epsilon}^{\pm}$, and thus  ${\bf p}=\pm i {\bf m}$   \cite{nietoheli}. Therefore at K1: ${\bf p}\cdot{\bf m}
^*=\pm i|\alpha_e|^2$. However at K2:  $\alpha_m=-\alpha^*_e$ and hence ${\bf p}\cdot{\bf m}^*=\mp i\alpha_e^2$.


\bibitem{comment ROT} In this regard we note that a reactive optical theorem in anisotropic media was reported  by  E. A. Marengo, A New Theory of the Generalized Optical Theorem in Anisotropic Media, IEEE Trans.  Antenn. Propag. {\bf 61},  2164-2179 (2013). 

\bibitem{re opteheli} In contrast with the energy OT and ROT,  both helicity OT and ROT  contain a real part in the extinction term. This is due to the  different functional form in the real and imaginary parts, $\bm{\mathcal F}$ and $\bm{\mathcal F}_{{\qui}}$, of the  complex flow of helicity $\bm{\mathcal F}_C$, as seen in Eqs. (\ref{helflow})-(\ref{bcpoy63}).

 Eq. (\ref{bcpoy63}) shows, at difference with those of  the CPV.

\bibitem{comment Vinfty} To prove that $\int_{\partial V_{\infty}}d^2 r \bm{\mathcal F}_{{\qui}}^{(s)}\cdot {\bf n}=0$, we  write the scattered field at points  in the far-zone  characterized by the direction unit vector ${\bf n}$ as: ${\bf E}^{(s)}({\bf n})={\bf e}({\bf n}) \exp(ikr)/r$ and 
${\bf H}^{(s)}({\bf n})={\bf h}({\bf n}) \exp(ikr)/r$,  ${\bf h}={\bf n}{\times\bf e}$. Then it is straightforward to see that $\int_{\partial V_{\infty}}d^2 r Im\{{\bf h}^*({\bf n})\times{\bf h}({\bf n})\}\cdot {\bf n}=
\int_{\partial V_{\infty}}d^2 r Im\{{\bf e}^*({\bf n})\times{\bf e}({\bf n})\}\cdot {\bf n}$. This latter equality and  the definition (\ref{reactflow}) of $ \bm{\mathcal F}_{{\qui}}^{(s)}$ constitute the proof.

\bibitem{barron}L. D. Barron, {\it  Molecular light scattering and optical activity}. Cambridge University Press,  (Cambridge, 2004).
University Press
\bibitem{schellman}J. A. Schellman,  Circular dichroism and optical rotation. Chem. Rev. {\bf 75}, 323–331  (1975).

\bibitem{comment2} In this connection, it should be noted that an expression akin to such dissymmetry factor, was  written in \cite{bliokh2} as what the authors called {\it  relative magnetoelectric response} of the particle to the illumination. This was done on employing what they named  {\it  magnetoelectric absorption rate},  determined  from $\Re [{\bf E}^{(i)}\cdot{\bf B}^{(i)*\,}]$, which they called {\it  magnetoelectric energy}; even though the physical process that produces it  was not reported. Here we have demonstrated that the  left side of Eqs. (\ref{reacdicroi1}) and (\ref{re_opteheliinfty}), as well as  (\ref{re_opteheli2}), describe   the mechanism through which this quantity appears and may be observed.

\bibitem{fortu}  M. F. Picardi, A. V. Zayats, and F. J. Rodríguez-Fortuño, Janus and Huygens Dipoles: Near-Field Directionality Beyond Spin-Momentum Locking, Phys. Rev. Lett. {\bf 120}, 117402  (2018).

\color{red}

\bibitem{calda} A. Krasnok, M. Caldarola, N. Bonod and Andrea Al\'{u}, Spectroscopy and Biosensing with Optically Resonant Dielectric Nanostructures, Advanced Optical Materials  1701094  (2018). DOI: 10.1002/adom.201701094.

\bibitem{zograf} G. P. Zograf, M. I. Petrov, S. V. Makarov, and Y. S. Kivshar,  All-dielectric thermonanophotonics,
Advances in Optics and Photonics  {\bf 13}, 643 (2021). 

\color{black}

\end{thebibliography}
\end{document}